\documentclass[
    reprint,
    prd,
    nofootinbib,
    amsmath,amssymb,
    aps,
    floatfix,
]{revtex4-2}
 
\usepackage{hyperref}
\usepackage{graphicx}
\usepackage{subcaption}
\usepackage{epigraph}
\usepackage{printlen}
\usepackage{tensor}
\usepackage[dvipsnames]{xcolor}
\newdimen\imageheight
\newdimen\imageheightsecond
\begin{document}


\title{The relativistic tidal tensor: general solutions for stationary axisymmetric spacetimes and the Hills mass of naked singularities}


\author{Wenkang \surname{Xin}}
\email[E-mail: ]{wenkang.xin@exeter.ox.ac.uk}
\affiliation{Rudolf Peierls Centre for Theoretical Physics, Beecroft Building, Parks Road, Oxford OX1 3PU, UK}
\author{Andrew \surname{Mummery}}
\email[E-mail: ]{amummery@ias.edu}
\affiliation{School of Natural Sciences, Institute for Advanced Study, 1 Einstein Drive, Princeton, NJ 08540, USA}
\affiliation{Rudolf Peierls Centre for Theoretical Physics, Beecroft Building, Parks Road, Oxford OX1 3PU, UK}
\date{\today}


\date{\today}

\begin{abstract}
    The tidal forces experienced on an orbit contain, in principle, information about the underlying spacetime an object is moving through. Astronomical observations often probe the properties of tidal forces in the relativistic regime, and could thus in principle be leveraged to examine the properties of strong-field gravity, provided that a general procedure for computing the relativistic tidal tensor is known. Existing techniques for deriving the tidal tensor rely on cumbersome, case-by-case methods. This paper introduces a unified analytical approach to deriving the tidal accelerations experienced by a test particle in any stationary, axisymmetric spacetime. This technique uses standard relativistic frame transformations and is built around the zero angular momentum observer frame. The method's utility is demonstrated in the four traditional black hole metrics: Schwarzschild, Reissner-Nordström, Kerr, and Kerr-Newman, as well as a particular wormhole metric. As an example of a possible astronomical application of this work, we discuss the concept of the Hills mass, the maximum mass at which a black hole can disrupt a star, and extend its definition to various naked singularity metrics.
\end{abstract}


\maketitle







\section{Introduction}

The differential evolution of nearby geodesics moving through curved spacetimes produce tidal accelerations, a directly observable consequence of the gravitational field. Many astrophysical observations probe the physics of tidal accelerations in the strong gravity regime, including so-called ``tidal disruption events'' (TDEs), and in principle could be leveraged to probe the detailed physics of strong field gravity. 

On rare occasion (typically once every $\sim 10^4-10^5$ years in a given galaxy \cite{Magorrian99}) a star may be perturbed by N-body gravitational interactions onto a near radial orbit about the central mass in a galaxy (naturally assumed to be a supermassive black hole). If the tidal force experienced by the star on its orbit overcomes the star's own self-gravity, then the star will become tidally unbound and subsequently accreted, a process which produces bright electromagnetic signatures \cite{Rees88}. A large number ($\sim 100$) of these so-called ``tidal disruption events'' have been discovered in recent years at observed wavelengths across the entire electromagnetic spectrum, from X-rays \citep[e.g.,][]{Greiner00, Cenko12, Guolo24, Grotova25}, and optical \citep[e.g.,][]{Gezari08, vanVelzen20, Hammerstein23, Yao23, Mummery_et_al_2024} to radio frequencies \citep[e.g.,][]{Alexander16, Goodwin22, Goodwin25}. 

One interesting consequence of tidal forces in black hole (BH) spacetimes is that their magnitudes are a decreasing function of BH mass when evaluated at the event horizon $(a_T \sim GM_\bullet R_\star/r^3 \sim 1/M_\bullet^2$ for $r\sim GM_\bullet/c^2)$. A consequence of this dropping tidal force is that there is a maximum mass (known as the Hills \cite{Hills1975} mass) at which a BH can tidally disrupt a star outside of its horizon and therefore produce potentially observable emission. 

The precise limiting value of the mass at which a BH can disrupt a star depends on the value of its spin parameter, a result that has been used in the TDE literature to constrain the spins of high-mass TDE systems \citep[e.g.,][]{Kesden12, Mummery2023, Mummery_et_al_2024}. However, in principle this framework can be extended to more general spacetimes, and rather than seeking to constrain the BH spin, one can seek to place bounds on deviations from simple Kerr-like metrics in the strong-field gravity regime. A similar (in spirit) idea can be applied to deviations from BH metrics and their impacts on the size and shape of the ``photon ring'' observed in EHT images of M87 \citep[e.g.,][]{EHT19, EHT2021}. 

To compute these limiting compact object masses, one must have a general procedure for computing tidal forces in (suitably general) relativistic spacetimes. This is complicated by the fact that relativistic tidal forces are only meaningful when the tidal tensor is computed in the local freely falling (LFF) frame of a test particle (see \cite{Pirani2009} for a discussion). Traditionally, this requires constructing a parallel-transported tetrad along the geodesic of a test particle. These so-called Fermi normal coordinates \cite{Manasse1963} have been used extensively to understand relativistic tidal forces \cite{Ishii2005,Cheng2013,Banerjee2018}, but their construction is generally a difficult task unless one restricts to spherical spacetimes or simple cases such as radial geodesics. A pioneering work by Marck \cite{Marck1983} gave the analytical expressions for the tidal tensor in the Kerr spacetime. Marck's result has been used in various works in the TDE community, including the stellar response to tidal forces \cite{Ivanov2001}, the relativistic Hills mass \cite{Mummery2023}, and the rate of TDEs \cite{Kesden2012}.

Extending Marck's computations to a different metric is cumbersome since every metric must be considered on a case-by-case basis. It is the purpose of this paper to present a unified analytical approach that is also computationally much simpler, so that tidal accelerations can be computed in a wide range of potentially interesting metrics. This approach borrows techniques from the black hole accretion community \citep{Thorne1974, Dauser2013, Ingram19}, using the zero angular momentum observers (ZAMO) frame \cite{Bardeen72} as an intermediary and standard special relativistic frame transformations. 

The paper is organised as follows. Section \ref{sec:motivation} discusses some astrophysical motivations for developing such a general framework. Section \ref{sec:theory} outlines the theoretical framework, and demonstrates how to compute the local tidal tensor and its properties in a general axisymmetric metric. Sections \ref{sec:spherical}, \ref{sec:kerr}, and \ref{sec:kn} are dedicated to the study of tidal forces in the traditional black hole metrics (Schwarzschild, Reissner-Nordström, Kerr and Kerr-Newman), presenting both new and known results. An exotic wormhole metric is analysed in Section \ref{sec:wormhole} as an additional example. We conclude with a discussion of potential future directions in Section \ref{sec:conclusion}. The appendices offer additional technical results, including the general form of the tidal tenor for any spherically symmetric spacetime.

We use natural units $G = c = 1$ and the signature $(-, +, +, +)$. Plain Greek indices such as $\mu, \nu, \rho$ are used for the distant observer frame (to be defined shortly); Roman indices in parentheses $(a), (b), (c)$ are for the zero angular momentum observer frame; while indices with tildes $\tilde{\alpha}, \tilde{\beta}, \tilde{\gamma}$ are for the local freely falling frame. Our discussion is restricted to neutral massive test particles following time-like geodesics.

\section{Astrophysical Motivation} \label{sec:motivation}
Before we discuss our general framework for computing tidal accelerations in relativistic spacetimes, we discuss our astrophysical motivation: using observations of TDEs to place constraints on the nature of high-curvature spacetimes. 

\subsection{Tidal forces and the Hills mass} \label{sec:astro_tidal_forces}
Consider Newtonian gravity and the tidal forces experienced by a star of radius $R_{\star}$ moving through a central potential sourced by some mass $M$. The tidal acceleration experienced by this star is roughly:
\begin{equation}\label{tidal_simple}
    a_{T} \sim \frac{GM}{r^{3}} R_{\star} \, .
\end{equation}
For a TDE to occur, this stretching tidal acceleration must overcome the attracting gravitational self-acceleration of the star \cite{Rees88}. The Newtonian self-gravity of a star is roughly:
\begin{equation}\label{self}
    a_{\text{self}}\sim \frac{GM_{\star}}{R_{\star}^{2}} \, ,
\end{equation}
where $M_{\star}$ is the mass of the star.

Combining Equations \ref{tidal_simple} and \ref{self} gives a rough estimate of the tidal radius, the orbital distance where tidal disruption occurs:
\begin{equation}
    r_{T} \sim \left( \frac{M}{M_{\star}} \right)^{1/3} R_{\star}  \, .
    \label{eq:tidal_radius}
\end{equation}
{In deriving this ``canonical'' tidal radius we have clearly neglected various order unity factors, which are conventionally brought together into a single factor. One can parametrise this correction in many (functionally equivalent) ways, but as tidal forces can be computed exactly we choose to parametrise the correction in terms of the deviation from the stellar self gravity from the simple estimate above,  defining a parameter $\eta$ by }
\begin{equation}
    \eta \equiv {a_{\rm self} R_\star^2 \over G M_\star}. 
\end{equation}
{Most analyses assume that $\eta = 1$ \citep[e.g.][]{Kesden12, MumBalb20b}, but this is definitely a simplification. If the star is rotating, for example, it is less bound and more easily disrupted $\eta < 1$. On the other hand, $\eta = 1$  might overestimate the ease at which stars are fully disrupted since it is possible that a star may lose its less bound outer layers while maintaining its more dense core \citep{Ryu20}, in which case $\eta > 1$ is relevant for a full disruption \citep[e.g.,][]{Phinney89}.
For a given metric and star, $\eta$ can be calibrated from numerical hydrodynamical simulations \citep[e.g.,][]{Guillochon13}, and is generally expected to be well approximated by the stellar core/average density ratio  $\eta \approx \rho_{\star, \rm core}/ \rho_{\star, \rm avg}$.  }

{Irrespective of these details regarding the stellar structure,} $r_{T}$ scales as $M^{1/3}$, and for a BH the horizon scales as $M$. For sufficiently massive BHs the tidal radius must eventually become smaller than the event horizon, meaning that TDEs will produce no observable emission. This highlights the crucial point that there exists a maximum mass of a BH that allows for an observationally relevant TDE to occur. This upper limit is known as the Hills mass \cite{Hills1975}, and can be derived by setting the tidal radius equal to the gravitational radius $(r_{T}\sim GM/c^{2})$, or:
\begin{equation}
    M_{\text{Hills}} \sim M_{0} \equiv \frac{R_{\star}^{3/2} c^3}{\eta^{1/2} G^{3/2} M_{\star}^{1/2}} \, ,
\end{equation}
where we use the subscript $0$ to indicate that $M_{0}$ is the dimensional scale derived from Newtonian gravity.

For a solar-type star this is roughly equal to the value $M_{0} \simeq 1.7 \times 10^{8} M_\odot$, and is therefore of a scale relevant for supermassive black holes in our Universe. The point is that while the dimensional factor $R_{\star}^{3/2} c^{3} G^{-3/2} M_{\star}^{-1/2}$ will not vary, there is, in a more careful theory, a dimensionless prefactor $f$ which depends on the detailed physics of the spacetime the star is moving through. 

In principle this dimensionless factor $f$ is a function only of the metric that describes the (relativistic) spacetime around compact objects{, and the internal stellar structure (via $\eta$)}. As we move into the era of large wide-field optical surveys and the $\sim 1000$'s of TDEs which they are expected to find \cite{Bricman20}, one can hope to place ``constraints on $f$'', and therefore on the nature of relativistic spacetimes, by determining the typical mass scale at which the observed distribution of TDEs rapidly drops to zero (i.e., the Hills mass where stars begin to be swallowed whole rather than disrupted outside of the event horizon). This framework is in many ways similar to using the radius of the shadow observed by the event horizon telescope to constrain more exotic deviations from typical black hole spacetimes (e.g., \cite{EHT2021} among many others).

More explicitly, we will be computing eigenvalues of the relativistic tidal tensor, which in this paper will be dimensionless and denoted as $\left\lvert \lambda \right\rvert$. In reality, $\lambda$ is measured in units of $c^6 G^{-2} M^{-2}$, where $M$ is the mass of the central object. Therefore, the force balance relationship discussed above can be expressed as:
\begin{equation}
    \left\lvert \lambda \right\rvert \frac{c^{6}}{G^{2}M^{2}} R_{\star} > \eta \frac{GM_{\star}}{R_{\star}^{2}} \, ,
\end{equation}
which can be solved for an upper bound of $M$, leading to an expression of the Hills mass:
\begin{equation}
    M_{\text{Hills}} = \left( \left\lvert \lambda \right\rvert_{\text{max}} \frac{c^{6} R_{\star}^{3}}{\eta G^{3} M_{\star}} \right)^{1/2} = \left\lvert \lambda \right\rvert_{\text{max}}^{1/2} M_{0} \, .
    \label{eq:hills_mass}
\end{equation}

We see that determining the maximum mass of a compact object which can disrupt a star becomes a problem of determining the largest possible tidal eigenvalue $\left\lvert \lambda \right\rvert_{\text{max}}$ that can be experienced by a test particle without plunging into a horizon or singularity{, and calibrating $\eta$ from hydrodynamical simulations. In this work we focus on the first of these two key steps, determining the largest tidal eigenvalue for as general as possible a class of metrics.} This requires answering two questions: how close can an object get to the centre $(r=0)$ of the spacetime while still being observable at infinity? And what is the tidal force experienced at this point of closest approach? Clearly, for this to be practical as a way of constraining strong-field gravity, one needs a general (and computationally simple) method of constructing relativistic tidal accelerations.

\section{Theoretical Framework} \label{sec:theory}

In a curved spacetime two geodesics that start close together will have a separation which in general either grows or shrinks as they traverse their respective trajectories. It is this ``geodesic deviation'' which describes the linear tidal response in a relativistic spacetime. Consider two geodesics separated by an (initially) infinitesimal displacement vector $\delta x^{\mu}$. We write them as $X^{\mu}(\tau)$ and $X'^{\mu}(\tau) = X^{\mu}(\tau) + \delta x^{\mu}(\tau)$, where $\tau$ is the proper time along the geodesic. To first order in $\delta x^{\mu}$, the separation evolves according to the geodesic deviation equation (see e.g., \cite[Chapter 7]{Hobson2006}):
\begin{equation}
    \frac{\mathrm{D}^{2}}{\mathrm{D} \tau^{2}} \delta x^{\mu} = \tensor{R}{^{\mu}_{\alpha \nu \beta}} U^{\alpha} U^{\beta} \delta x^{\nu} \, .
    \label{eq:geodesic_deviation}
\end{equation}
Here, $U^\mu$ is the 4-velocity of a test particle (with respect to a coordinate system which we are yet to specify). The operator $\mathrm{D}/\mathrm{D}\tau$ denotes the covariant derivative along the geodesic with respect to proper time, and is explicitly given by ${\mathrm{D}}/{\mathrm{D}}\tau = U^{\alpha} \nabla_{\alpha}$, where $\nabla_\alpha$ is the usual covariant derivative.

The tidal tensor is therefore defined as:
\begin{equation}
    \tensor{C}{^{\mu}_{\nu}} \equiv -\tensor{R}{^{\mu}_{\alpha \nu \beta}} U^{\alpha} U^{\beta} \, ,
    \label{eq:tidal_tensor}
\end{equation}
so that we can write:
\begin{equation}
    \frac{\mathrm{D}^{2}}{\mathrm{D} \tau^{2}} \delta x^{\mu} = -\tensor{C}{^{\mu}_{\nu}} \delta x^{\nu} \, ,
\end{equation}
in analogy with Newtonian tidal forces. 

Equation \ref{eq:geodesic_deviation} describes the deviation of geodesics, as measured according to some coordinate system which we are yet to specify. The relevant coordinate system for computing \textit{physical} tidal accelerations are those that describe the local freely falling frame of the test particle, as this is the ``tidal acceleration'' which the particle itself would measure. 

The traditional approach to computing physical tidal forces is to explicitly solve for a coordinate tetrad that is parallel-transported along the geodesic (i.e., explicitly finding a coordinate description which represents the free-falling frame of the particle at each point along its geodesic). With this tetrad known, one can derive the local tidal tensor. However, constructing such a suitable parallel-transported tetrad is a challenging task that can typically only be done on a case-by-case basis. Notably, Marck \cite{Marck1983} developed a general set of parallel-transported tetrads for the Kerr metric. For radial geodesics in spherical spacetimes, a simple tetrad can be constructed for various metrics \citep[e.g.,][]{Crispino2016, Sharif2018, Hong2020, Vandeev2021, Vandeev2022, Lima2022, Arora2024}.

It is the purpose of this work to introduce a more computationally straightforward and general approach to deriving the local tidal tensor, using a different set of relativistic frame transformations which can be applied generally to a wide class of metrics. Instead of explicitly constructing a parallel-transported tetrad, we derive the form of the tidal tensor in a simpler coordinate system and then transform it to the LFF frame via standard relativistic frame transformations. This must always be possible because $\tensor{C}{^{\mu}_{\nu}}$ is a good rank-2 tensor (in the sense that it is constructed via the contraction of the Riemann tensor and 4-velocities) and therefore can be transformed between frames.

As we shall show, this procedure bypasses the need for explicitly computing the parallel-transported tetrad, the main computational difficulty in the traditional approach. We step through our procedure in the following sub-sections. 

\subsection{ZAMO transformation of tidal tensor} \label{sec:zamo}

Following the approach first put forward by \cite{Bardeen1972}, a general stationary, axisymmetric spacetime has a metric which can always be written in the form:
\begin{equation}
    \mathrm{d}s^{2} = -e^{2\nu} \mathrm{d}t^{2} + e^{2\psi} \left( \mathrm{d}\phi - \omega \mathrm{d}t \right)^{2} + e^{2\mu_{1}} \mathrm{d}r^{2} + e^{2\mu_{2}} \mathrm{d}\theta^{2} \, ,
    \label{eq:metric}
\end{equation}
where $\nu$, $\psi$, $\mu_{1}$, $\mu_{2}$, and $\omega$ are functions of $r$ and $\theta$ only. We henceforth refer to this frame as the ``distant observer frame'', as typically $t, r, \theta$ and $\phi$ are defined with respect to some observer at $r\to\infty$ (although this may not always be the case). 

A so-called ``zero angular momentum observer''\footnote{This frame is sometimes referred to as the ``locally non-rotating frame'' (LNRF) in the literature.} (ZAMO) is described by the 4-position $r = \text{const}$, $\theta = \text{const}$, and $\phi = \omega t + \text{const}$. The orthonormal tetrad carried by the ZAMO observer is known in full generality, and is given by\footnote{ ``$\mathbf{e}$'' in tetrads is not to be confused with ``$e$'' as the natural logarithm base.}:
\begin{equation}
    \begin{split}
        \tensor{\mathbf{e}}{^\mu_{(t)}} &= e^{-\nu} \left( \frac{\partial }{\partial t} + \omega \frac{\partial }{\partial \phi} \right) \, , \\
        \tensor{\mathbf{e}}{^\mu_{(r)}} &= e^{-\mu_{1}} \frac{\partial }{\partial r} \, , \\
        \tensor{\mathbf{e}}{^\mu_{(\theta)}} &= e^{-\mu_{2}} \frac{\partial }{\partial \theta} \, , \\
        \tensor{\mathbf{e}}{^\mu_{(\phi)}} &= e^{-\psi} \frac{\partial }{\partial \phi} \, .
    \end{split}
\end{equation}
The corresponding dual basis (covariant basis) is:
\begin{equation}
    \begin{split}
        \tensor{\mathbf{e}}{_{\mu}^{(t)}} &= e^{\nu} \mathrm{d}t \, , \\
        \tensor{\mathbf{e}}{_{\mu}^{(r)}} &= e^{\mu_{1}} \mathrm{d}r \, , \\
        \tensor{\mathbf{e}}{_{\mu}^{(\theta)}} &= e^{\mu_{2}} \mathrm{d}\theta \, , \\
        \tensor{\mathbf{e}}{_{\mu}^{(\phi)}} &= e^{\psi} \left( -\omega\mathrm{d}t + \mathrm{d}\phi \right) \, .
    \end{split}
\end{equation}
One can use this tetrad to project any tensor into the ZAMO frame, a transformation that corresponds to describing physical quantities by their components as measured by the local ZAMO observer. Crucially, it is simple to verify by explicit computation that 
\begin{equation}
    \tensor{g}{_{(a)(b)}} = \tensor{\mathbf{e}}{^\mu_{(a)}} \tensor{\mathbf{e}}{^\nu_{(b)}} g_{\mu\nu} = \eta_{(a)(b)} = \mathrm{diag}(-1,1,1,1) \, ,
\end{equation}
meaning that one can use the local transformation laws of special relativity in this frame (as the ZAMO observer infers a flat spacetime locally). 

With the ZAMO tetrad defined, a test particle's 4-velocity $U^{\mu}$ in the distant observer frame $(t, r, \theta, \phi)$ can be projected onto the ZAMO basis:
\begin{equation}
    U^{(a)} = U^{\mu} \tensor{\mathbf{e}}{_{\mu}^{(a)}} \, ,
\end{equation}
where $U^{(a)}$ is the 4-velocity of the particle measured in the ZAMO frame.

The Riemann tensor in the distant observer frame can also be transformed to the ZAMO frame:
\begin{equation}
    \tensor{R}{^{(a)}_{(b)(c)(d)}} = \tensor{\mathbf{e}}{_{\mu}^{(a)}} \tensor{\mathbf{e}}{^{\nu}_{(b)}} \tensor{\mathbf{e}}{^{\rho}_{(c)}} \tensor{\mathbf{e}}{^{\sigma}_{(d)}} \tensor{R}{^{\mu}_{\nu \rho \sigma}} \, .
\end{equation}
Since we place the ZAMO at the test particle's position and spacetime is locally Minkowski in the ZAMO frame, a simple Lorentz boost yields the Riemann tensor in the LFF frame (i.e., the physically relevant frame) \cite{Bardeen1972}:
\begin{equation}
    \tensor{R}{^{\tilde{\mu}}_{\tilde{\alpha} \tilde{\nu} \tilde{\beta}}} = \tensor{\varLambda}{_{(a)}^{\tilde{\mu}}} \tensor{\varLambda}{^{(b)}_{\tilde{\alpha}}} \tensor{\varLambda}{^{(c)}_{\tilde{\nu}}} \tensor{\varLambda}{^{(d)}_{\tilde{\beta}}} \tensor{R}{^{(a)}_{(b)(c)(d)}} \, ,
\end{equation}
where the Lorentz boost matrix can be explicitly written as:
\begin{equation}
    \begin{split}
        &\tensor{\varLambda}{^{\mu}_{\nu}} \equiv \\
        &\mathbb{I}_{4} + \begin{bmatrix}
            \gamma - 1            & -\gamma \beta^{r}                                      & -\gamma\beta^{\theta}                                     & -\gamma\beta^{\phi}                                       \\
            -\gamma\beta^{r}      & (\gamma - 1) \frac{(\beta^{r})^{2}}{\beta^{2}}         & (\gamma - 1) \frac{\beta^{r}\beta^{\theta}}{\beta^{2}}    & (\gamma - 1) \frac{\beta^{r}\beta^{\phi}}{\beta^{2}}      \\
            -\gamma\beta^{\theta} & (\gamma - 1) \frac{\beta^{\theta}\beta^{r}}{\beta^{2}} & (\gamma - 1) \frac{(\beta^{\theta})^{2}}{\beta^{2}}       & (\gamma - 1) \frac{\beta^{\theta}\beta^{\phi}}{\beta^{2}} \\
            -\gamma\beta^{\phi}   & (\gamma - 1) \frac{\beta^{\phi}\beta^{r}}{\beta^{2}}   & (\gamma - 1) \frac{\beta^{\phi}\beta^{\theta}}{\beta^{2}} & (\gamma - 1) \frac{(\beta^{\phi})^{2}}{\beta^{2}}
        \end{bmatrix} \, ,
    \end{split}
\end{equation}
where $\mathbb{I}_{4}$ is the $4 \times 4$ identity matrix $\mathbb{I}_{4}={\rm diag}(1, 1, 1, 1)$ and:
\begin{equation}
    \begin{split}
        \gamma &\equiv \frac{1}{\sqrt{1 - \beta^{2}}} \, , \\
        \beta &\equiv \left[ {(\beta^{r})^{2} + (\beta^{\theta})^{2} + (\beta^{\phi})^{2}} \right]^{1/2} \, , \\
        \beta^{i} &\equiv U^{(i)} / U^{(t)} \quad \text{for} \quad i = r, \theta, \phi \, .
    \end{split}
\end{equation}
With the Riemann tensor now known in the LFF frame, the physical local tidal tensor can be constructed:
\begin{equation}
    \tensor{C}{^{\tilde{\mu}}_{\tilde{\nu}}} = -\tensor{R}{^{\tilde{\mu}}_{\tilde{\alpha} \tilde{\nu} \tilde{\beta}}} U^{\tilde{\alpha}} U^{\tilde{\beta}} \, ,
    \label{eq:local_tidal_tensor}
\end{equation}
where $U^{\tilde{\alpha}} = (1, 0, 0, 0)$ is the 4-velocity of the particle in its own LFF frame.

The above steps provide a general algorithm for deriving the local tidal tensor using relativistic frame transformations. Given a stationary, axisymmetric metric, one can always compute the Riemann tensor in the distant observer frame $(t, r, \theta, \phi)$, transform it to the ZAMO frame using the known ZAMO tetrad, project the 4-velocity of the particle into the ZAMO, and then Lorentz boost to the LFF frame and construct the local tidal tensor. It is worth mentioning that a somewhat similar but less general approach was used in \cite{Chicone2006} for the Kerr metric.

Once the 4-position and 4-velocity of a test particle are specified, the tidal tensor $\tensor{C}{^{\tilde{\mu}}_{\tilde{\nu}}}$ can always be explicitly, and rather simply, evaluated. Obtaining the 4-velocities for computing $\tensor{C}{^{\tilde{\mu}}_{\tilde{\nu}}}$ of course requires integrating the geodesic equations, either analytically or numerically. Finding an analytical solution in general requires the  of various first integrals of the geodesic equations, but this is not always true in complete generality. Numerical integration is always available as a last resort.

\subsection{Properties of local tidal tensor}

The local tidal tensor $\tensor{C}{^{\tilde{\mu}}_{\tilde{\nu}}}$ can be viewed as a $4 \times 4$ matrix encoding the physical tidal accelerations experienced by a test particle. To find the accelerations, one must diagonalise $\tensor{C}{^{\tilde{\mu}}_{\tilde{\nu}}}$ to obtain the eigenvalues and eigenvectors. A negative (positive) eigenvalue indicates a stretching (compressing) force along the corresponding eigenvector, as can be inferred by inspecting Equation \ref{eq:geodesic_deviation}.

In general, the tidal tensor always has a zero eigenvalue in the time direction. To prove this, consider the 4-velocity of the particle $U^{\tilde{\alpha}} = (1, 0, 0, 0)$ as observed in its LFF frame. This implies that non-zero components of the local tidal tensor in Equation \ref{eq:local_tidal_tensor} must satisfy $\tilde{\alpha} = \tilde{\beta} = 0$:
\begin{equation}
    \tensor{C}{^{\tilde{\mu}}_{\tilde{\nu}}} = -\tensor{R}{^{\tilde{\mu}}_{0\tilde{\nu}0}} \, .
\end{equation}
From the symmetry of the Riemann tensor, we have $\tensor{R}{^{\tilde{\mu}}_{000}} = -\tensor{R}{^{\tilde{\mu}}_{000}} = 0$. Therefore, $\tensor{C}{^{0}_{\tilde{\mu}}} = \tensor{C}{^{\tilde{\mu}}_{0}} = 0$ for all $\tilde{\mu}$, so the local tidal tensor can be viewed as a singular matrix with a zero determinant. Since the determinant is a scalar invariant, the conclusion applies to $\tensor{C}{^{\mu}_{\nu}}$ in the distant observer frame as well.

This is also the reason why it is customary to take the tidal tensor as a $3 \times 3$ matrix by neglecting the temporal components, as is done in some literature. For consistency, we still present the tidal tensor as a $4 \times 4$ matrix with its first row and first column filled with zeros.

Crucially, a tidal tensor is \textit{not} traceless in general as the trace depends on the Ricci tensor:
\begin{equation}
    \begin{split}
        \tensor{C}{^{\mu}_{\mu}} &= -\tensor{R}{^{\mu}_{\alpha \mu \beta}} U^{\alpha} U^{\beta} \\
        &= -R_{\alpha \beta} U^{\alpha} U^{\beta} \\
        &= -\left( T_{\alpha \beta} - \frac{1}{2} g_{\alpha \beta} T \right) U^{\alpha} U^{\beta} \, ,
    \end{split}
\end{equation}
where the final step uses the Einstein field equations.

The tidal tensor is traceless for Ricci-flat metrics (such as Kerr) but not for a general metric. As we explicitly show in the subsequent sections, the trace is non-zero in Reissner-Nordström and Kerr-Newman spacetimes, where the Ricci tensor does not vanish due to a non-zero electromagnetic stress-energy tensor present in all regions of spacetime. Since the trace is a scalar invariant, the conclusion applies to $\tensor{C}{^{\tilde{\mu}}_{\tilde{\nu}}}$ in the LFF frame as well.

\subsection{Eigenvectors and eigenvalues}

The three non-zero eigenvalues of the matrix $\tensor{C}{^{\tilde{\mu}}_{\tilde{\nu}}}$ represent the physical tidal accelerations experienced in the LFF frame of the particle. It is these eigenvalues that set the scale of the tidal acceleration, but the directions are determined by the eigenvectors, which we shall denote as $x^{\tilde{\nu}}_{[i]}$, where $i = 1, 2, 3$ denotes the three eigenvectors. These eigenvectors contain information regarding the direction of the tidal acceleration in the LFF frame of the particle. This may not always be the most useful frame in which to know the direction of the acceleration as, for example, one may be simulating the tidal accelerations in a fixed coordinate system (e.g., Boyer-Lindquist coordinates). 

To transform the eigenvectors from the LFF frame back to the distant observer frame one simply inverts the process above, or explicitly:
\begin{equation}
    x_{[i]}^\mu = \tensor{\mathbf{e}}{^\mu_{(a)}} \tensor{\left(\Lambda^{-1}\right)}{^{(a)}_{\tilde \nu}} x^{\tilde \nu}_{[i]} \, ,
\end{equation}
where $\Lambda^{-1}$ is identical to $\Lambda$ but with $\beta^{i} \to -\beta^{i}$ everywhere.

\section{Spherically Symmetric Spacetimes} \label{sec:spherical}

Let us begin with the spherically symmetric case, a subset of axisymmetric spacetimes that admit a ZAMO. In terms of the general metric functions in Equation \ref{eq:metric}, a stationary, spherically symmetric spacetime can be described by:
\begin{equation} \label{eq:spherical_metric}
    \begin{split}
        &e^{2\nu} = A(r) \, , \qquad e^{2\psi} = r^{2} \sin^{2}{\theta} \, , \\
        &e^{2\mu_{1}} = B(r) \, , \qquad e^{2\mu_{2}} = r^{2} \, , \\
        &\omega = 0 \, ,
    \end{split}
\end{equation}
where $A(r)$ and $B(r)$ are prescribed functions of $r$ that depend on the specific spacetime.

As a result of the spherical symmetry of the problem, we may set $\theta = \pi/2$ and $\dot{\theta} = 0$ for any orbit without loss of generality. The first integrals to the geodesic equations of a test particle are (see e.g., \cite{Chandrasekhar1998}):
\begin{equation}
    \begin{split}
        \frac{\mathrm{d}t}{\mathrm{d}\tau} &= \frac{\epsilon}{A} \, , \\
        B\left( \frac{\mathrm{d}r}{\mathrm{d}\tau} \right)^{2} &= R(r) \equiv \frac{\epsilon^{2}}{A} - \frac{l_{z}^{2}}{r^{2}} - 1 \, , \\
        \frac{\mathrm{d}\theta}{\mathrm{d}\tau} &= 0 \, , \\
        \frac{\mathrm{d}\phi}{\mathrm{d}\tau} &= \frac{l_{z}}{r^{2}} \, .
    \end{split}
    \label{eq:spherical_geodesics}
\end{equation}
Here, we have used following constants of motion:
\begin{equation}
    E \equiv -p_{t} \, , \qquad L \equiv p_{\phi} \, , \qquad m^{2} \equiv -g_{\mu \nu} p^{\mu} p^{\nu} \, ,
\end{equation}
which are the energy, angular momentum and rest mass of the particle respectively.

We scaled the conserved quantities with the rest mass $m$ so that we can take $m = 1$:
\begin{equation}
    \epsilon \equiv \frac{E}{m} \, , \qquad l_{z} \equiv \frac{L}{m} \, .
\end{equation}
This above definition encompasses a broad class of spherical metrics. Tidal tensors in specific instances have been previously studied in the literature \citep[e.g.,][]{Crispino2016, Sharif2018, Hong2020, Vandeev2021, Vandeev2022, Lima2022, Arora2024}, many of which focus on particles in radial trajectories (i.e., $\dot{\phi} = 0$). However, while letting $\dot{\theta} = 0$ incurs no loss of generality, setting $\dot{\phi} = 0$ is a strong and restrictive assumption: the test particle must have zero angular momentum.

Our ZAMO method yields a fully general set of expressions for the local tidal tensor in spherical spacetimes without any assumptions on $\epsilon$ or $l_{z}$. This result is presented in full in Appendix \ref{sec:app}. Although lengthy, the algebraic expressions remain tractable functions of $A$, $B$ and their derivatives. Their full form has been verified to reproduce the works cited above when specialised to each metric.

Let us take the Reissner-Nordström (RN) spacetime as an illustrative example, whose metric functions are:
\begin{equation}
    A(r) = \frac{1}{B(r)} = 1 - \frac{2M}{r} + \frac{Q^{2}}{r^{2}} \, .
\end{equation}
In this expression $M$ is the mass of the central object (which can be a black hole or naked singularity depending on the relative magnitude of $Q/M$) and $Q$ is a measure of its charge. We will set $M = 1$ without loss of generality.

If $Q^{2} \leq M^{2}$, the metric describes a BH with horizons at $r_{\pm} = M \pm \sqrt{M^{2} - Q^{2}}$; if $Q^{2} > M^{2}$, the metric describes a naked singularity (NS). One recovers the Schwarzschild metric by setting $Q = 0$.

For radial geodesics ($\theta = \pi/2$ and $\dot{\theta} = \dot{\phi} = 0$), previous work \cite{Crispino2016} has shown that the local Riemann tensor has only a few non-zero components: $\tensor{R}{^{\tilde{r}}_{\tilde{t}\tilde{r}\tilde{t}}} = A''(r)/2$ and $\tensor{R}{^{\tilde{\theta}}_{\tilde{t}\tilde{\theta}\tilde{t}}} = \tensor{R}{^{\tilde{\phi}}_{\tilde{t}\tilde{\phi}\tilde{t}}} = A'(r)/2r$. This results in a diagonal local tidal tensor that is straightforward to analyse.

For radial geodesics in the RN spacetime, the local tidal tensor can be explicitly derived and simplified using the ZAMO method:
\begin{equation}
    C =
    \frac{1}{r^{4}}
    \begin{bmatrix}
        0 & 0           & 0         & 0         \\
        0 & 3Q^{2} - 2r & 0         & 0         \\
        0 & 0           & r - Q^{2} & 0         \\
        0 & 0           & 0         & r - Q^{2}
    \end{bmatrix} \, ,
\end{equation}
which agrees with the results in \cite{Crispino2016}.

Lifting the condition $\dot{\phi} = 0$ or $l_{z} = 0$ significantly complicates the tidal tensor. However, the ZAMO method still allows for analytical solutions of its eigenvalues:
\begin{equation}
    \begin{split}
        \lambda_{1} &= \frac{r^{3} - Q^{2} r^{2}}{r^{6}} \, , \\
        \lambda_{2} &= \frac{-2r^{3} + 3Q^{2} r^{2} - 3l_{z}^{2} r + 4Q^{2}l_{z}^{2}}{r^{6}} \, , \\
        \lambda_{3} &= \frac{r^{3} - Q^{2} r^{2} + 3l_{z}^{2} r - 2Q^{2}l_{z}^{2}}{r^{6}} \, .
    \end{split}
    \label{eq:rn_eigenvalues}
\end{equation}
These expressions readily reduce to the $l_{z} = 0$ case: $\lambda_{2}$ corresponds to the radial direction, while $\lambda_{1}$ and $\lambda_{3}$ correspond to the angular directions. Setting $Q = 0$ yields the well-known Schwarzschild tidal accelerations; taking $r \to \infty$ yields the Newtonian results. Notice that the eigenvalues do not depend on the energy $\epsilon$ of the particle. This turns out to be a fairly common feature of spherical spacetimes.

Notably, each of the three eigenvalues vanishes at some positive radius. $\lambda_{1}$ has a trivial root at $r=Q^{2}$. For $\lambda_{2}$ and $\lambda_{3}$, their numerators are cubic polynomials with the sign sequence $(-, +, -, +)$ and $(+, -, +, -)$ respectively. Both have exactly three sign changes, implying at least one positive root by Descartes' rule of signs. This is similar to the situation with $l_{z} = 0$ \cite{Crispino2016}.

The sum of the eigenvalues gives the trace of the tidal tensor:
\begin{equation}
    \sum_{i=1}^{3} \lambda_{i} = \frac{Q^{2}(r^{2} + 2l_{z}^{2})}{r^{6}} \, .
    \label{eq:rn_trace}
\end{equation}
The trace is strictly positive for $Q \neq 0$, indicating that the overall tidal effect of $Q$ in the RN spacetime is compressive. This is a result of a non-vanishing stress-energy tensor due to the presence of charge that produces electromagnetic energy density everywhere.

\subsection{Hills mass in the RN metric}

\subsubsection{RN black holes} \label{sec:rn_ibco}

For $Q^2 \leq 1$, the RN metric describes a black hole. In this situation the radius of closest approach (i.e., the relevant radius for computing the Hills mass) is that of the innermost bound circular orbit (IBCO). For stars that start at rest from spatial infinity (a good approximation for a star in the galactic potential), the ISCO describes the \textit{separatrix} between orbits that plunge into the horizon and those that return to infinity \cite{Levin2008, MummeryBalbus2023PRD}. The IBCO therefore represents the closest a star could get to a RN BH (starting at rest from spatial infinity) without crossing the horizon \cite{Mummery2023}. The IBCO radius is defined from the radial part in the first integrals of motion (Equation \ref{eq:spherical_geodesics}). Indeed, the IBCO exists at the location $r=\chi$ where $R(r = \chi) = R'(r = \chi) = 0$ and $\epsilon = 1$ \cite{Levin2008, MummeryBalbus2023PRD}. Substituting the RN metric functions leads to a condition on the angular momentum:
\begin{equation}
    l_{z}^{2} = \frac{2\chi^{3} - Q^{2}\chi^{2}}{Q^{2} - 2\chi + \chi^{2}} \, ,
\end{equation}
and a characteristic equation for $\chi$:
\begin{equation}
    \chi^{3} - 4\chi^{2} + 4Q^{2}\chi - Q^{4} = 0 \, .
\end{equation}
For a given $Q$, the characteristic equation may have multiple positive roots. The largest one is taken as the IBCO radius, as the particle cannot reach the inner solutions on an IBCO geodesic.

\begin{figure}[ht!]
    \centering
    \includegraphics[width=\columnwidth]{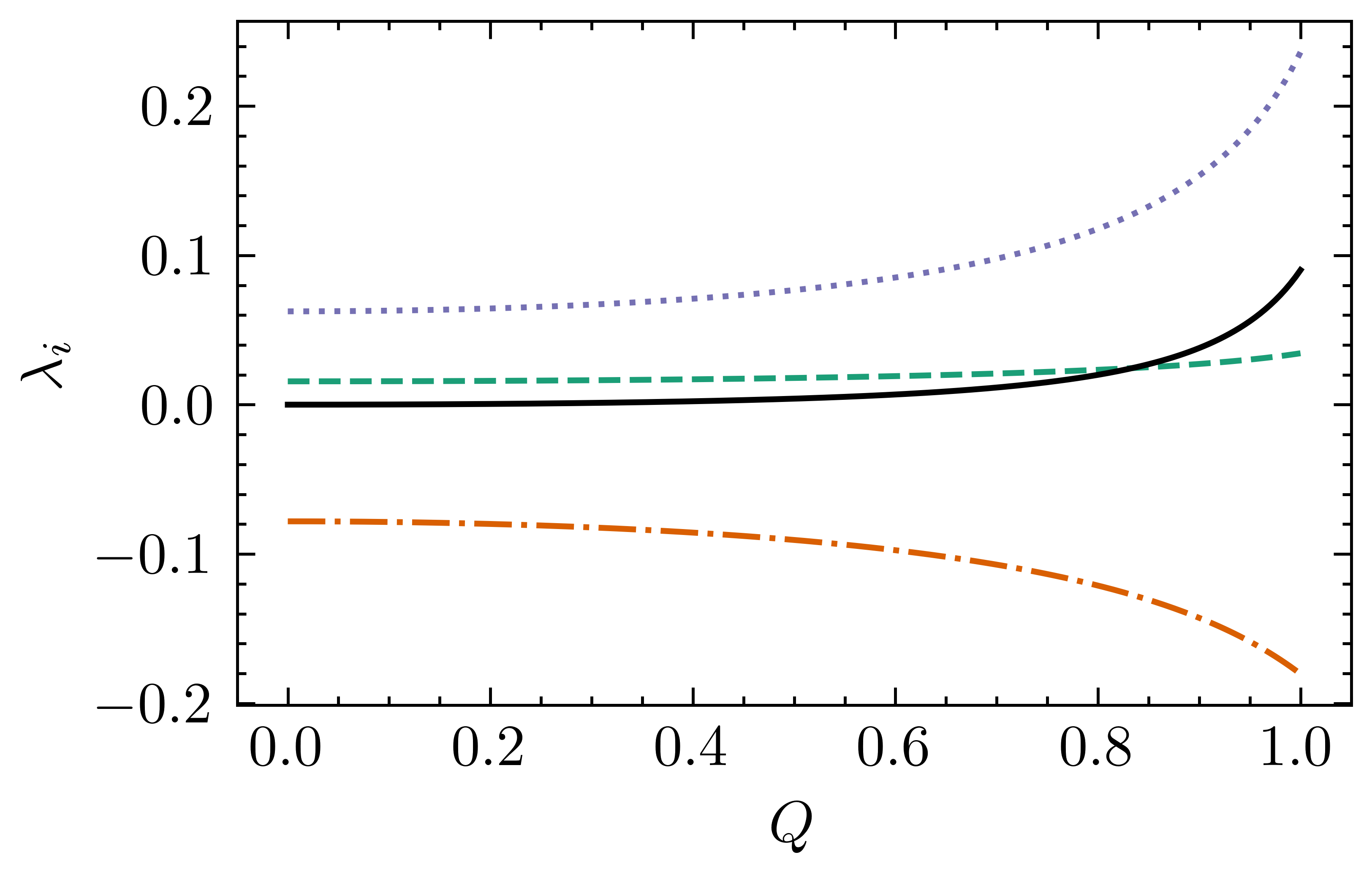}
    \caption{Tidal eigenvalues on IBCO in RN spacetime, with $\lambda_{1}$ (green dashed), $\lambda_{2}$ (orange dash-dotted), $\lambda_{3}$ (purple dotted), and the sum $\sum \lambda_{i}$ (black solid).}
    \label{fig:ibco_eigenvalues_rn}
\end{figure}

Given the expressions of $\chi$ and $l_{z}$, the tidal eigenvalues (Equation \ref{eq:rn_eigenvalues}) depend only on $Q$. We plot the eigenvalues against $Q \in [0, 1]$ in Figure \ref{fig:ibco_eigenvalues_rn} and remind the reader that they are measured in units of $c^{6}G^{-2}M^{-2}$. Observe that their sum (trace of the tidal tensor) increases with $Q$, consistent with the earlier point that tidal forces are net compressive in the RN spacetime.

We reiterate that IBCOs are of interest because they represent the closest meaningful approach to the BH and thus the largest tidal forces. That is, Figure \ref{fig:ibco_eigenvalues_rn} shows the maximum tidal accelerations that can be experienced by a unit-energy particle without falling into the horizon.

The values of the tidal accelerations at the end points of the curves can be determined exactly. At $Q = 0$, one recovers the Schwarzschild results: $\chi = 4$ with $\lambda_{1} = 1/64$, $\lambda_{2} = -5/64$, and $\lambda_{3} = 4/64$. At $Q = 1$, the characteristic equation is solved to give the largest positive solution of $\chi = (3 + \sqrt{5})/2 \approx 2.618$, leading to $\lambda_{1} = (13\sqrt{5} - 29)/2 \approx 0.0344$, $\lambda_{2} = -(5\sqrt{5} - 11) \approx -0.180$, and $\lambda_{3} = \sqrt{5} - 2 \approx 0.236$.

Now consider the stretching (negative) eigenvalue $\lambda_{2}$ as this is the relevant eigenvalue for disruption. We can express its magnitude as a function of $Q$ and note that $\left\lvert \lambda_{2} \right\rvert(Q)$ is a monotonically increasing function for $Q \in [0, 1]$ with $\left\lvert \lambda_{2} \right\rvert(0) = 5/64$ and $\left\lvert \lambda_{2} \right\rvert(1) = 5\sqrt{5} - 11$. This means that a RN BH will always have a larger Hills mass than a Schwarzschild (uncharged) one.

Since the tidal eigenvalues have units of $c^{6}G^{-2}M^{-2}$, disruption occurs when the tidal acceleration exceeds the self-gravitational acceleration of the star:
\begin{equation}
    \left\lvert \lambda_{2} \right\rvert \frac{c^{6}}{G^{2} M^{2}} \gtrsim \frac{G M_{\star}}{R_{\star}^{3}} \, ,
\end{equation}
or equivalently:
\begin{equation}
    \left\lvert \lambda_{2} \right\rvert \gtrsim \frac{G^{3} M_{\star} M^{2}}{c^{6} R_{\star}^{3}} \, .
\end{equation}
This means that the Hills mass of a BH in RN spacetime is the upper limit of $M$ satisfying the above inequality:
\begin{equation}
    M_{\text{Hills}} = \left( \left\lvert \lambda_{2} \right\rvert \frac{c^{6} R_{\star}^{3}}{\eta G^{3} M_{\star}} \right)^{1/2} = \left\lvert \lambda_{2} \right\rvert^{1/2} M_{0} \, ,
\end{equation}
where we have used the Newtonian mass scale $M_{0}$.

Setting $Q = 0$ gives the Schwarzschild Hills mass \cite{Hills75, Kesden12, Mummery2023}:
\begin{equation}
    M_{\text{Hills}}^{\text{Schw}} = \frac{\sqrt{5}}{8} M_{0} \, ,
\end{equation}
while setting $Q = 1$ gives the Hills mass of an extremal RN BH:
\begin{equation}
    M_{\text{Hills}}^{\text{RN max}} = (5\sqrt{5} - 11)^{1/2} M_{0} \, ,
\end{equation}
which is about $1.52$ times the Schwarzschild Hills mass.

\subsubsection{RN naked singularities}

For $Q^{2} > 1$, we have a NS without horizons. As explained in Section \ref{sec:motivation}, we consider the closest approach orbits, which in this case are those followed by a unit-energy particle with zero angular momentum $l_{z}$. This is because in the NS regime, the issue of plunging into a horizon does not arise. Intuitively, particles with no angular momentum follow head-on trajectories directly towards the origin. As long as the particle does not reach the origin, it can still escape to infinity and produce observable emission. We therefore set $\epsilon = 1$ and $l_{z} = 0$ and the radial potential $R(r)$ simplifies to:
\begin{equation}
    R(r) = \frac{2r - Q^{2}}{r^{2} - 2r + Q^{2}} \, ,
\end{equation}
which has a root at $r = Q^{2}/2$, where head-on particles are repelled. This implies that the gravitational effect of $Q$ is repulsive, a phenomenon noted in previous literature \citep[e.g.,][]{Pugliese2011, Patil2012}.

At this closest approach, the tidal eigenvalues are:
\begin{equation}
    \lambda_{1} = -\frac{8}{Q^{6}} \, , \qquad \lambda_{2} = \frac{32}{Q^{6}} \, , \qquad \lambda_{3} = -\frac{8}{Q^{6}} \, .
    \label{eq:rn_ns_eigenvalues}
\end{equation}
Note that $\lambda_{2}$ is now positive (compressing), while $\lambda_{1}$ and $\lambda_{3}$ are negative (stretching); contrast this to the BH case, where the IBCO eigenvalues have the opposite signs. The magnitudes are significantly larger than the BH regime due to the much smaller radial distance.

We can similarly derive the Hills mass for naked singularities of the RN metric. Using either $\lambda_{1}$ or $\lambda_{3}$ from Equation \ref{eq:rn_ns_eigenvalues}, disruption occurs when:
\begin{equation}
    \frac{8}{Q^{6}} \frac{c^{6}}{G^{2} M^{2}} > \eta \frac{G M_{\star}}{R_{\star}^{3}} \, ,
\end{equation}
which leads to a Hills mass of:
\begin{equation}
    M_{\text{Hills}}^{\text{RN NS}} = \left[ \frac{8 c^{6} R_{\star}^{3}}{\eta G^{3} M_{\star} Q^{6}} \right]^{1/2} = \left( \frac{8}{Q^{6}} \right)^{1/2} M_{0} \, .
    \label{eq:rn_ns_hills_mass}
\end{equation}

For $Q$ just above $1$, the Hills mass of a RN NS is up to $8/\sqrt{5} \approx 3.58$ times the Schwarzschild Hills mass. However, as $Q$ increases further, the Hills mass decreases rapidly with $Q^{-3}$.

\section{Axisymmetric Spacetime --- Kerr} \label{sec:kerr}

Our attention is now directed to the more complicated case of Kerr spacetime, represented by the metric functions (see Equation \ref{eq:metric}):
\begin{equation}
    \begin{split}
        &e^{2\nu} = \frac{\Sigma \Delta}{\mathcal{A}} \, , \qquad e^{2\psi} = \sin^{2}{\theta} \frac{\mathcal{A}}{\Sigma} \, , \\
        &e^{2\mu_{1}} = \frac{\Sigma}{\Delta} \, , \qquad e^{2\mu_{2}} = \Sigma \, , \\
        &\omega = \frac{2Mra}{\mathcal{A}} \, ,
    \end{split}
\end{equation}
where:
\begin{equation}
    \begin{split}
        \Sigma &\equiv r^{2} + a^{2}\cos^{2}{\theta} \, , \\
        \Delta &\equiv r^{2} - 2Mr + a^{2} \, , \\
        \mathcal{A} &\equiv (r^{2} + a^{2})^{2} - \Delta a^{2} \sin^{2}{\theta} \, .
    \end{split}
\end{equation}
$M$ is the mass of the BH/NS and $a$ is the spin parameter that quantifies its angular momentum. If $a^{2} \le M^{2}$, the metric describes a BH with horizons at $r_{\pm} = M \pm \sqrt{M^{2} - a^{2}}$; if $a^{2} > M^{2}$, the metric describes a naked singularity. We will keep setting $M = 1$ without loss of generality.

First integrals of the geodesic equations in Kerr spacetime are well-known (see \cite{Bardeen1972, Carter2009} for a discussion), and can be expressed in terms of the following constants of motion:
\begin{equation}
    E \equiv -p_{t} \, , \qquad L \equiv p_{\phi} \, , \qquad m^{2} \equiv -g_{\mu \nu} p^{\mu} p^{\nu} \, ,
\end{equation}
which are the energy, angular momentum and rest mass of the particle respectively.

An additional conserved quantity called the Carter constant is defined as:
\begin{equation}
    \mathcal{Q} \equiv p_{\theta}^{2} + \cos^{2}{\theta} \left[ a^{2} (E^{2} - 1) + \frac{L^{2}}{\sin^{2}{\theta}} \right] \, .
\end{equation}
An alternative definition of the Carter constant proves more useful as its value is non-negative:
\begin{equation}
    \mathcal{K} \equiv \mathcal{Q} + (L - aE)^{2} \, .
\end{equation}
As before we scale the conserved quantities by the rest mass $m$ so that we can take $m = 1$:
\begin{equation}
    \epsilon \equiv \frac{E}{m} \, , \quad l_{z} \equiv \frac{L}{m} \, , \quad q \equiv \frac{\mathcal{Q}}{m^{2}} \, , \quad k \equiv \frac{\mathcal{K}}{m^{2}} \, .
\end{equation}
In the following analysis of orbital dynamics, we are primarily interested in particles with unit energy $\epsilon = 1$. It turns out that in this limit, $q$ represents the angular momentum of the particle projected on the equatorial plane:
\begin{equation}
    q = p_{\theta}^{2} + l_{z}^{2} \cot^{2}{\theta} = l_{x}^{2} + l_{y}^{2} \, .
\end{equation}
This means that $\sqrt{l_{z}^{2} + q}$ is the total angular momentum of the particle, which motivates the definition of the conserved variables $\Lambda$ and $\varphi$:
\begin{equation}
    \Lambda = \cos{\varphi} \equiv \frac{\sqrt{q}}{\sqrt{l_{z}^{2} + q}} \, .
    \label{eq:lambda}
\end{equation}
The equations of motion then become:
\begin{equation}
    \begin{split}
        \Sigma \frac{\mathrm{d}t}{\mathrm{d}\tau} &= -a(a \epsilon \sin^{2}{\theta} - l_{z}) + (r^{2} + a^{2}) \frac{T}{\Delta} \, , \\
        \Sigma^{2} \left( \frac{\mathrm{d}r}{\mathrm{d}\tau} \right)^{2} &= R(r) \, , \\
        \Sigma^{2} \left( \frac{\mathrm{d}\theta}{\mathrm{d}\tau} \right)^{2} &= \Theta(\theta) \, , \\
        \Sigma \frac{\mathrm{d}\phi}{\mathrm{d}\tau} &= -\left( a \epsilon - \frac{l_{z}}{\sin^{2}{\theta}} \right) + a \frac{T(r)}{\Delta} \, .
    \end{split}
    \label{eq:kerr_first_integrals}
\end{equation}
The functions $T$, $R$, and $\Theta$ are defined as:
\begin{equation}
    \begin{split}
        T(r) &\equiv \epsilon (r^{2} + a^{2}) - al_{z} \, , \\
        R(r) &\equiv T^{2} - \Delta (r^{2} + k) \, , \\
        \Theta(\theta) &\equiv k - (l_{z} - a\epsilon)^{2} - \cos^{2}{\theta} \left[ a^{2} (1 - \epsilon^{2}) + \frac{l_{z}^{2}}{\sin^{2}{\theta}} \right] \, .
    \end{split}
\end{equation} 
Note that in the Kerr (and Kerr-Newman) metric, a degeneracy exists in the choice of signs between $l_{z}$ and $a$, since the first integrals are invariant if both quantities switch signs: $l_{z} \to -l_{z}$ and $a \to -a$. We will restrict to $a \ge 0$ and allow $l_{z}$ to take any value; orbits with $l_{z} > 0$ are called prograde, while those with $l_{z} < 0$ are retrograde.

In the Kerr spacetime, the ZAMO algorithm {developed in this work provides a general (i.e., as a function of $r, \theta, \phi$) and completely analytic (written in terms of elementary functions) form of the local tidal tensor.} {The} expressions {describing this local tidal tensor are however}  extremely {long winded, and we do not report them directly here. They are instead included in a Mathematica notebook available with this manuscript}. 

{While lengthy, these analytical expressions mean} it is straightforward (and computationally fast) to conduct a {comprehensive} study of the tidal tensors by integrating the geodesic equations\footnote{See Appendix for details of the implementation.} {and using these new analytical expressions along the particle trajectory}. 

\begin{figure}[ht!]
    \centering
    \begin{subfigure}[b]{\columnwidth}
        \centering
        \includegraphics[width=\textwidth]{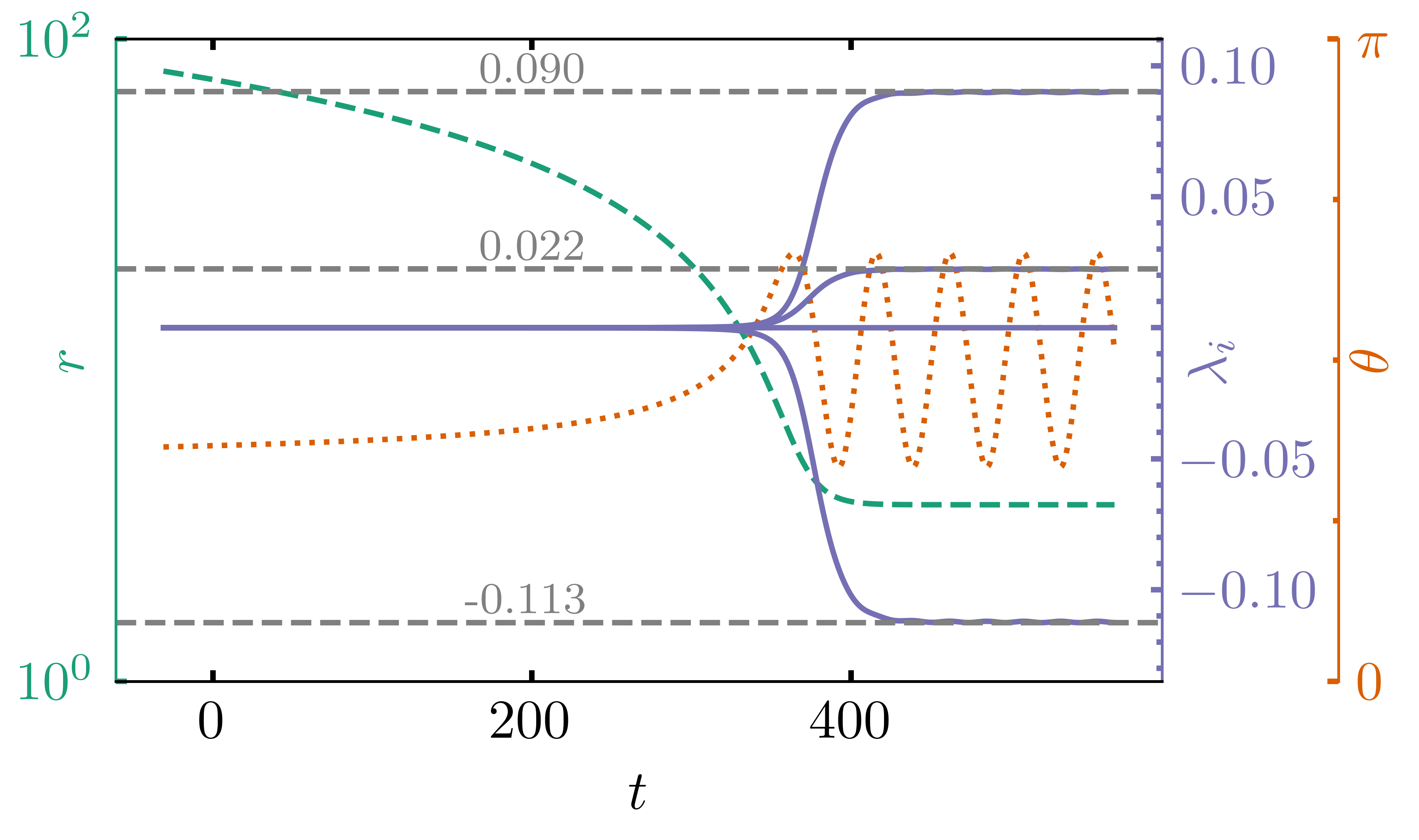}
        \caption{$\Lambda = 0.5$, $a = 0.25$}
    \end{subfigure}
    \begin{subfigure}[b]{\columnwidth}
        \centering
        \includegraphics[width=\textwidth]{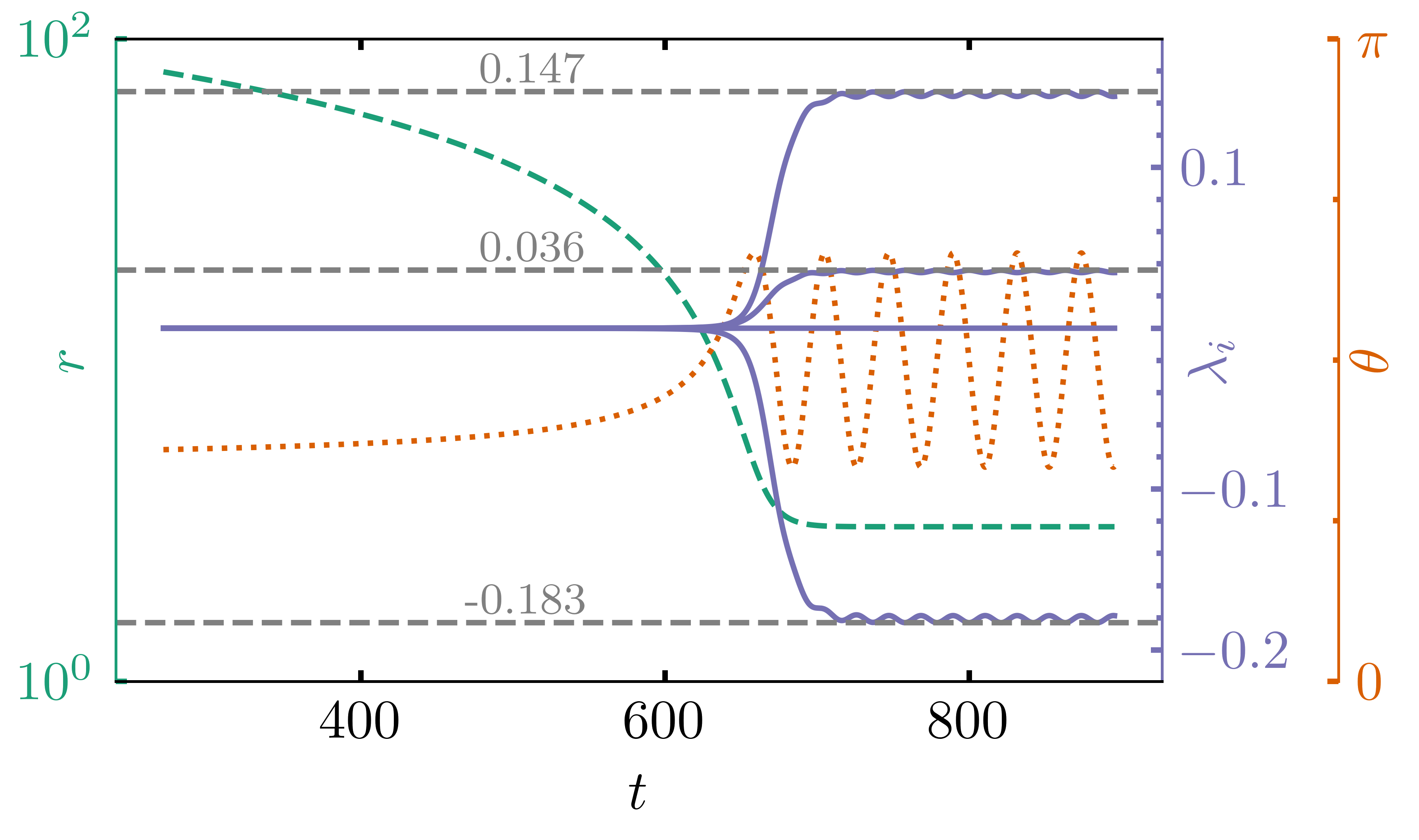}
        \caption{$\Lambda = 0.5$, $a = 0.50$}
    \end{subfigure}
    \begin{subfigure}[b]{\columnwidth}
        \centering
        \includegraphics[width=\textwidth]{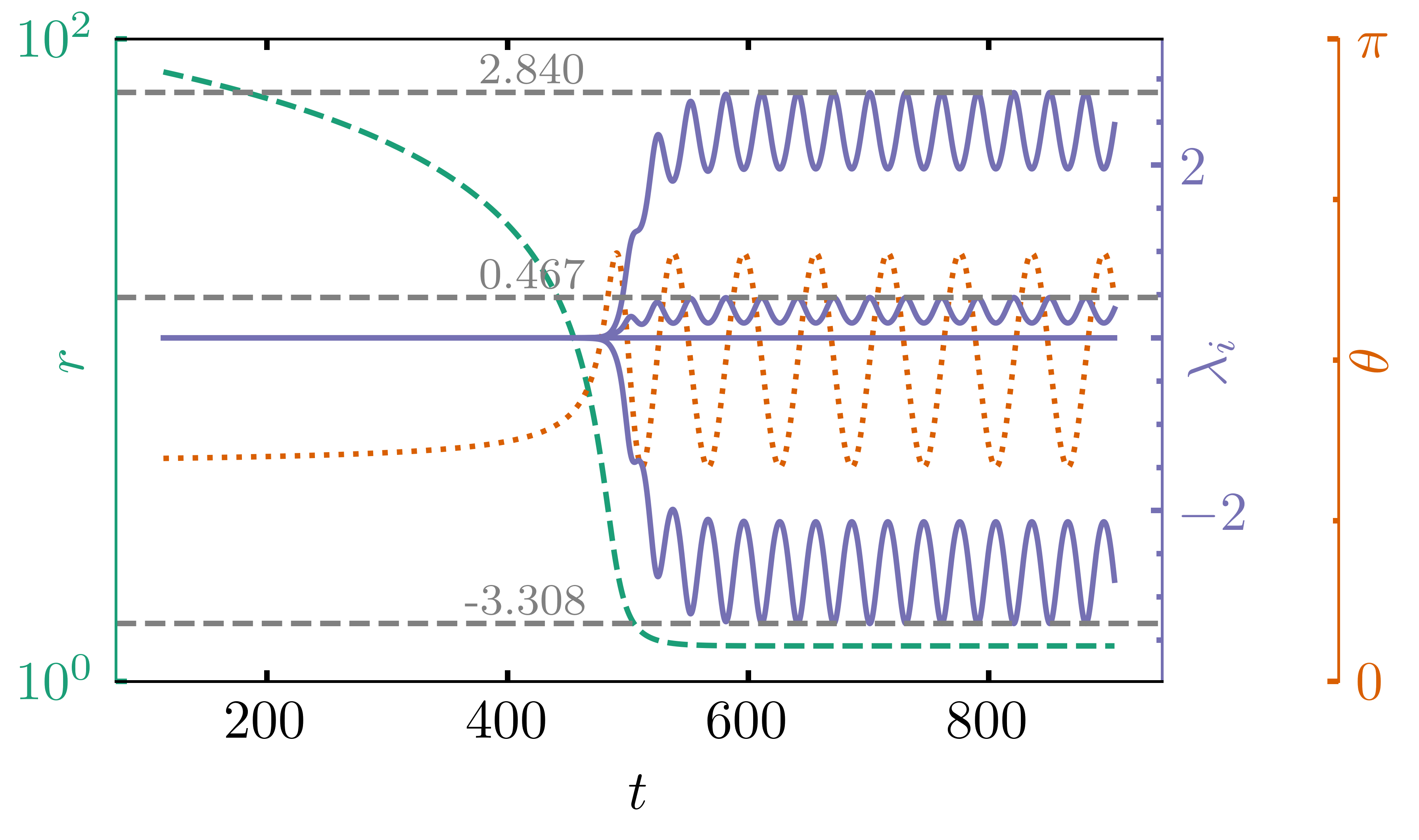}
        \caption{$\Lambda = 0.5$, $a = 0.99$}
    \end{subfigure}
    \caption{Evolution of tidal eigenvalues along prograde IBSO geodesics in Kerr BH, with radial distance (green dashed), polar angle (orange dotted), and tidal eigenvalues (purple solid). Grey dashed lines indicate theoretical eigenvalues as the particle crosses the equatorial plane.}
    \label{fig:kerr_simulations}
\end{figure}

In Figure \ref{fig:kerr_simulations}, we plot the evolution of coordinates and tidal eigenvalues for a particle on an innermost bound {\it spherical} orbit (IBSO; spherical because the particle is no longer confined to a plane, see e.g., \cite{Hod2013, Mummery2023}) trajectory. Similar to the previously explained IBCOs, the IBSO radius is the point at which $\epsilon = 1$ and $R(r = \chi) = R'(r = \chi) = 0$, leading to the twin constraints:
\begin{equation}
    l_{z} = \frac{1}{a(\chi - 1)} \left[ \chi^{2} - a^{2} - \chi^{1/2} \left( \chi^{2} - 2\chi + a^{2} \right) \right] \, ,
\end{equation}
and:
\begin{equation} \label{eq:ibso_kerr}
    \begin{split}
        0 &= 3\chi^{6} - 4 \chi^{5} + a^{2}(9\Lambda^{2} - 2) \chi^{4} \\
            &\quad+ a^{2} (4 - 16\Lambda^{2}) \chi^{3} \\
            &\quad+ a^{2} \left( -a^{2} + 6a^{2} \Lambda^{2} \right) \chi^{2} \\
            &\quad+ a^{6} \Lambda^{2} \\
            &\quad- 8 \chi^{7/2} \left( \chi^{2} - 2\chi + a^{2} \right) \, .
    \end{split}
\end{equation}
In Figure \ref{fig:kerr_simulations} a particle with $\epsilon = 1$ is sent from spatial infinity at an inclination angle $\cos^{-1}\Lambda$, where $\Lambda$ is defined in Equation \ref{eq:lambda}; the particle's angular momentum is set according to the above condition so that the particle approaches the IBSO radius asymptotically.

Several observations arise. As expected, there is always one zero tidal eigenvalue, while the non-zero eigenvalues have a magnitude which depends on both $r$ and $\theta$ (note that the latter has a much weaker effect). Local maxima in the magnitude of eigenvalues occur on the equatorial plane ($\theta = \pi/2$), where the effect of spin is most pronounced; this is consistent with studies that find the strongest tidal accelerations on the equatorial plane \cite{Kesden2012, Mummery2023}.

We may cross-reference the numerical results with results of Marck \cite{Marck1983}. On the equatorial plane ($\theta = \pi/2$), where the results of Marck simplify nicely, the local tidal tensor has the eigenvalues \cite{Kesden2012}:
\begin{equation}
    \begin{split}
        \lambda_{1} &= \frac{1}{r^{3}} \left( 1 + \frac{3k}{r^{2}} \right) \, , \\
        \lambda_{2} &= \frac{1}{r^{3}} \, , \\
        \lambda_{3} &= -\frac{1}{r^{3}} \left( 2 + \frac{3k}{r^{2}} \right) \, .
    \end{split}
\end{equation}
These expressions can be compared with the results from the ZAMO approach. Reference lines in Figure \ref{fig:kerr_simulations} are drawn by substituting the IBSO radius $r=\chi$ and the corresponding $k$ into the above results. We note that our numerical results asymptote to the analytical results as $t\to \infty$ and when $\theta$ crosses $\pi/2$, as they must.

\subsection{Hills mass of Kerr naked singularities}

We defer the discussion of Kerr BH Hills masses to Section \ref{sec:kn} and focus here on naked singularities. Consider a Kerr NS where $a^{2} > M^{2}$ and no horizon exists. As in treating a KN NS, we consider a unit-energy particle on zero angular momentum orbits to find the radius of closest approach. We therefore set $\epsilon = 1$ and $l_{z} = 0$ and the radial potential $R(r)$ simplifies to:
\begin{equation}
    R(r) = 2r^{3} + 2a^{2}r = 2r(r^{2} + a^{2}) \, .
\end{equation}
The radial potential has a single real root at $r = 0$, so that a head-on particle can reach the origin of a Kerr NS where the tidal forces diverge. This implies that, by having a small but non-zero angular momentum, a particle can get arbitrarily close to a Kerr NS, experience arbitrarily large tidal forces, and still escape to infinity. Therefore, a Kerr NS has an \textit{infinite} Hills mass. This is fundamentally different from the BH regime, where the meaningful tidal forces are limited by the non-zero IBSO radii, leading to a finite Hills mass. 

We see immediately that observations of a population of tidal disruption events will be capable of placing strong constraints on the prevalence of any Kerr-like naked singularities in our Universe. Indeed, there is an observed super-exponential cut-off in detected TDE emission above $M \sim 10^8 M_\odot$ seen in TDEs \cite{Yao23, MummeryVV25}. This can therefore likely be used to place tight constraints on the number density of Kerr-like naked singularities in our Universe (i.e., they cannot represent more than a fraction of $f\sim 1/N_{\rm TDE}$ of all massive compact objects, where $N_{\rm TDE} \sim 100$ is the number of known TDE systems).

\section{Axisymmetric Spacetime --- Kerr-Newman} \label{sec:kn}

We now examine the Kerr-Newman (KN) metric, the most general form of a black hole. The metric functions (see Equation \ref{eq:metric}) are similar to Kerr except for modifications of $\omega$ and $\Delta$ to include the charge parameter:
\begin{equation}
    \begin{split}
        &e^{2\nu} = \frac{\Sigma \Delta}{\mathcal{A}} \, , \qquad e^{2\psi} = \sin^{2}{\theta} \frac{\mathcal{A}}{\Sigma} \, , \\
        &e^{2\mu_{1}} = \frac{\Sigma}{\Delta} \, , \qquad e^{2\mu_{2}} = \Sigma \, , \\
        &\omega = \frac{(2Mr - Q^{2})a}{\mathcal{A}} \, ,
    \end{split}
\end{equation}
where:
\begin{equation}
    \begin{split}
        \Sigma &\equiv r^{2} + a^{2}\cos^{2}{\theta} \, , \\
        \Delta &\equiv r^{2} - 2Mr + a^{2} + Q^{2} \, , \\
        \mathcal{A} &\equiv (r^{2} + a^{2})^{2} - \Delta a^{2} \sin^{2}{\theta} \, .
    \end{split}
\end{equation}
$M$ is the mass of the BH/NS, $a$ is the spin parameter, and $Q$ is the charge parameter. If $a^{2} + Q^{2} \le M^{2}$, the metric describes a BH with horizons at $r_{\pm} = M \pm \sqrt{M^{2} - a^{2} - Q^{2}}$; if $a^{2} + Q^{2} > M^{2}$, the metric describes a naked singularity. We keep setting $M = 1$ without loss of generality.

The conserved quantities of the KN spacetime are the same as those of the Kerr spacetime. Similarly, the first integrals of the geodesics are the entirely analogous to those in Kerr once the charge parameter $Q$ is included in the functions $\omega$ and $\Delta$.

As with the  Kerr metric, the Kerr-Newman tidal tensor {can be derived in a full analytical (in terms of elementary functions) form, for orbits of arbitrary position $r, \theta, \phi$ (see notebook). Unsurprising, this tidal tensor } contains complicated expressions that resist {obvious} simplification, {and we do not report each component in this manuscript as they are lengthy}. However, it is still possible to derive the trace of the tidal tensor:
\begin{equation}
    \begin{split}
        \sum_{i=1}^{3} \lambda_{i} = \frac{Q^{2}}{\Sigma^{3}} (r^{2} + 2k - a^{2} \cos^{2}{\theta}) \, ,
    \end{split}
\end{equation}
where $k$ is the alternative Carter constant scaled by the particle's rest mass.

The trace is generally not zero unless $Q = 0$. In fact, upon substituting $\epsilon = 1$, which is the case for IBSOs, the trace becomes:
\begin{equation}
    \frac{Q^{2}}{\Sigma^{3}} \left[ r^{2} + 2 \left( \frac{l_{z}^{2}}{\tan^{2}{\theta}} + p_{\theta}^{2} \right) + 2(a - l_{z})^{2} - a^{2} \cos^{2}{\theta} \right] \, ,
\end{equation}
which is manifestly positive for $Q \neq 0$, indicating a compressive tidal effect.

Unlike Kerr, there is very little literature on the tidal forces by a KN BH. Part of the reason may be a lack of existing tetrad formulations like the one Marck derived for the Kerr spacetime. In fact, in the closing paragraph of his paper \cite{Marck1983}, Marck commented that his approach could be easily generalised to the KN spacetime. This turns out to be incorrect: Marck's method of calculating the tidal tensor relies on Ricci-flatness so that the Weyl tensor is identical to the Riemann tensor. This is not the case in the KN spacetime where the stress-energy tensor is non-zero.

\begin{figure}[ht!]
    \centering
    \begin{subfigure}[b]{\columnwidth}
        \centering
        \includegraphics[width=\textwidth]{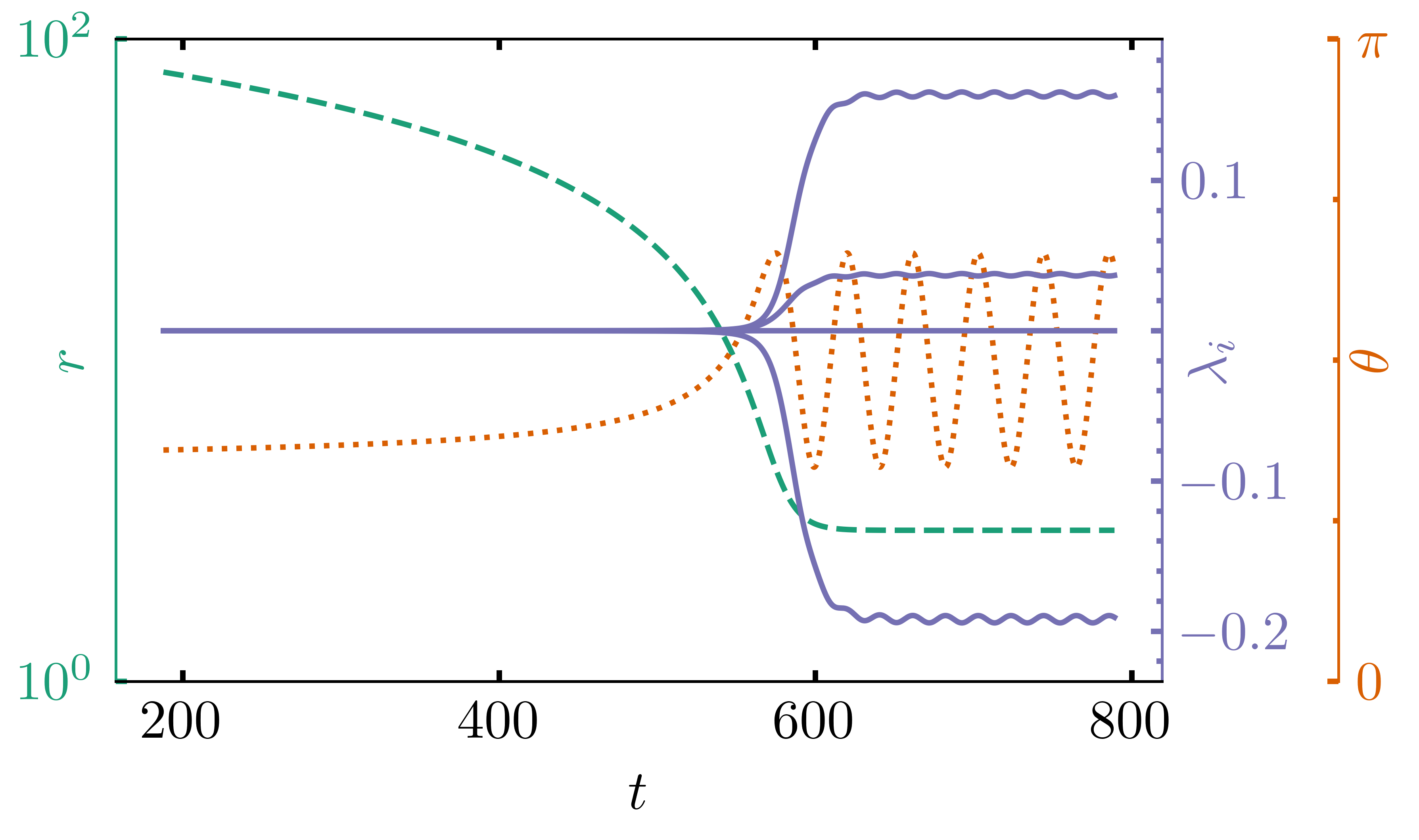}
        \caption{$\Lambda = 0.5$, $a = 0.5$, $Q = 0.25$}
    \end{subfigure}
    \begin{subfigure}[b]{\columnwidth}
        \centering
        \includegraphics[width=\textwidth]{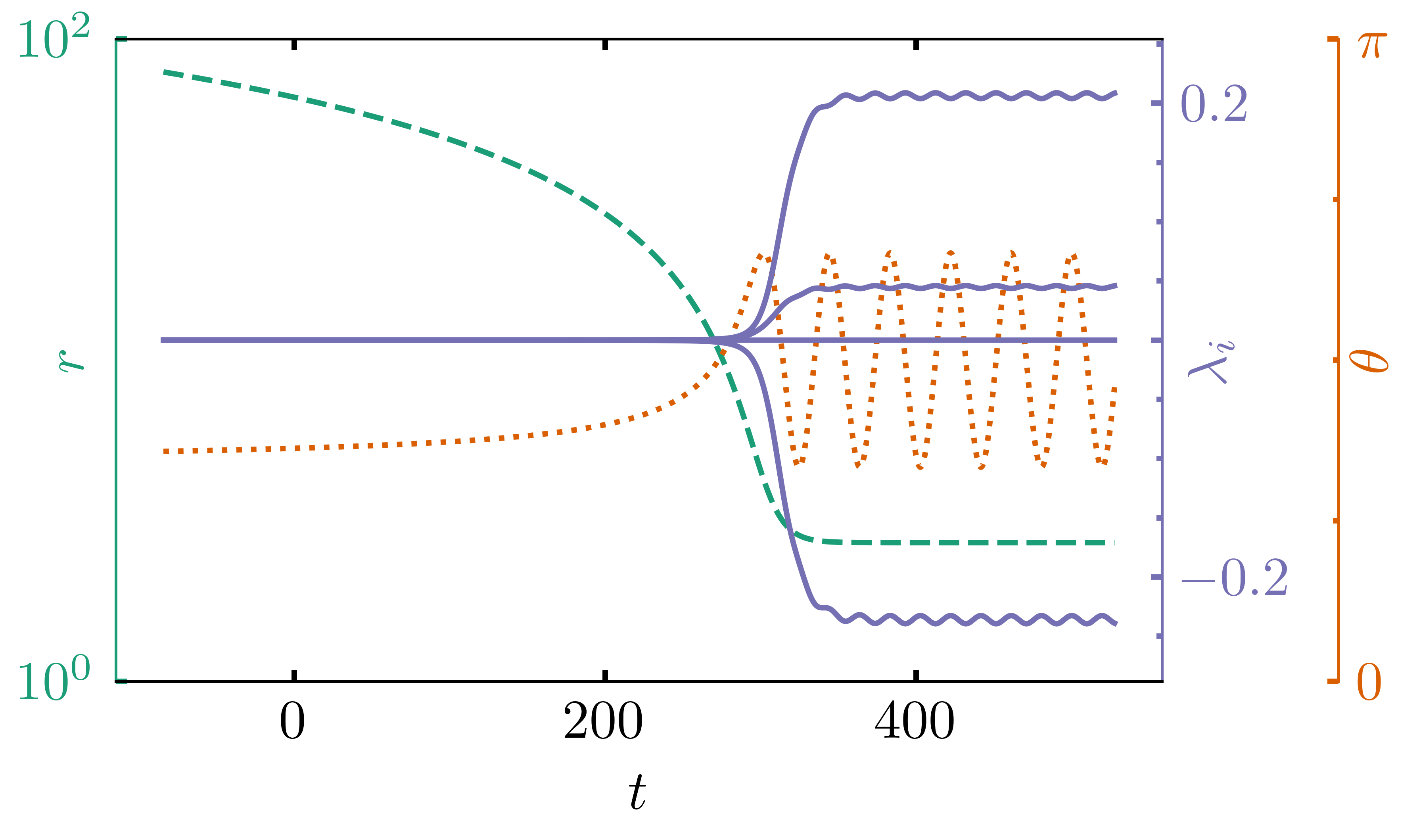}
        \caption{$\Lambda = 0.5$, $a = 0.5$, $Q = 0.5$}
    \end{subfigure}
    \begin{subfigure}[b]{\columnwidth}
        \centering
        \includegraphics[width=\textwidth]{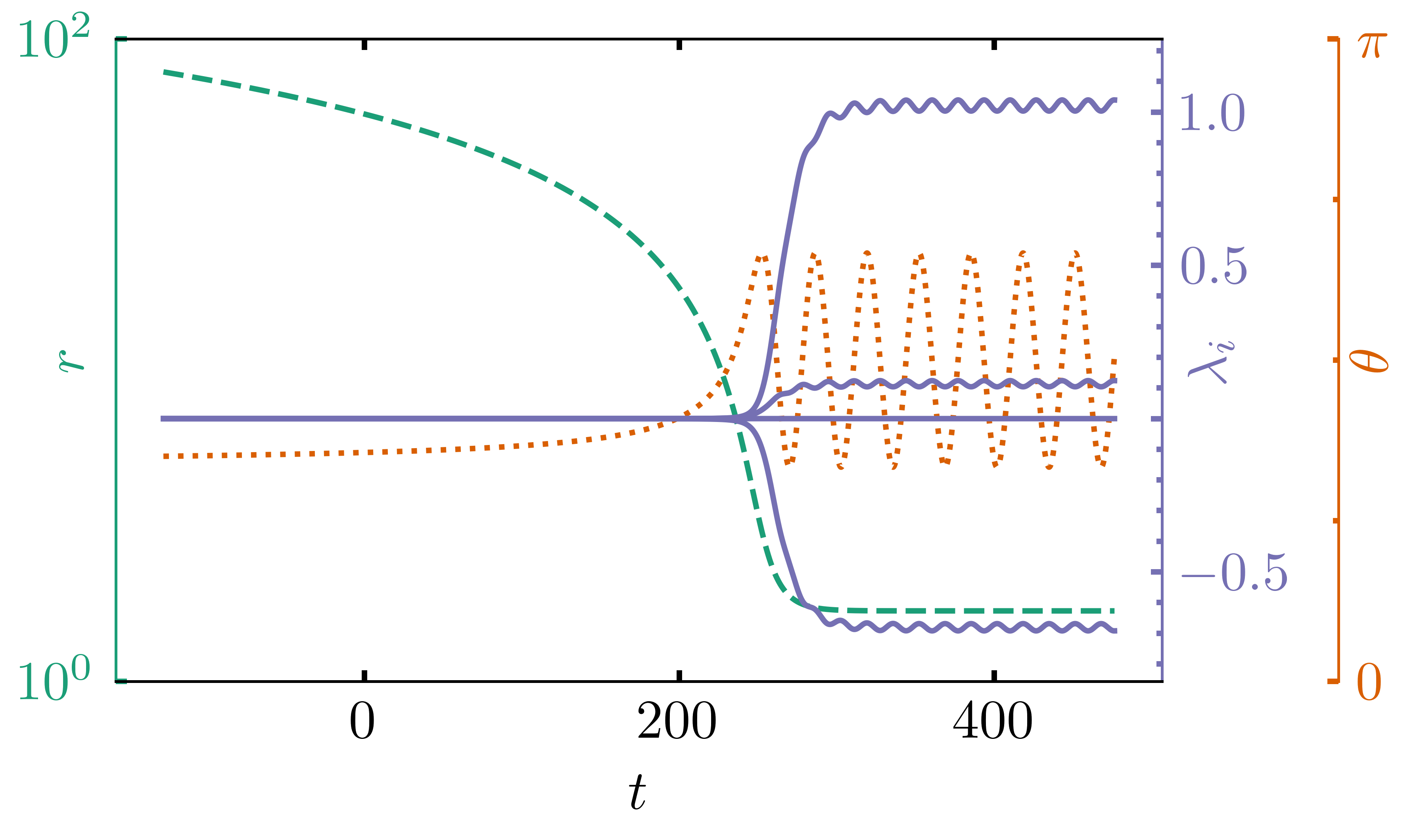}
        \caption{$\Lambda = 0.5$, $a = 0.5$, $Q = 0.85$}
    \end{subfigure}
    \caption{Evolution of tidal eigenvalues along prograde IBSO geodesics in KN BH, with radial distance (green dashed), polar angle (orange dotted), and tidal eigenvalues (purple solid).}
    \label{fig:kn_simulations}
\end{figure}

{In the following section we demonstrate the utility of our analytical tidal tensor.}  Particles are set on IBSOs (whose properties we discuss in the next subsection) and the evolution is plotted in Figure \ref{fig:kn_simulations}. The general behaviour of the tidal accelerations is similar to Kerr, but the presence of charge allows the particle to get closer to the BH and experience stronger tidal accelerations. The intuition of this observation is explained in the following discussion. Note that the strongest tidal accelerations occur on the equatorial plane, same as Kerr. For large $Q$, the strongest compressing (positive) eigenvalue can have a larger magnitude than the strongest stretching (negative) one. This is due to the presence of charge that has a net compressive tidal effect, similar to the RN spacetime.

\subsection{Hills mass in the KN metric}
\subsubsection{IBSO in KN}
Let us consider IBSOs in the KN spacetime with $a^{2} + Q^{2} \le 1$. Any results here carry over to the simpler Kerr spacetime by setting $Q = 0$. The KN spacetime has the same radial potential as Kerr defined in Equations \ref{eq:kerr_first_integrals}, except that $\Delta$ now has an additional term $Q^{2}$:
\begin{equation}
    \begin{split}
        R(r) &= \left[ \epsilon (r^{2} + a^{2}) - al_{z} \right]^{2} \\
        &\quad- \Delta \left[ r^{2} + (l_{z} - a\epsilon)^{2} + l_{z}^{2} \frac{\Lambda^{2}}{1 - \Lambda^{2}} \right] \, ,
    \end{split}
\end{equation}
where we have used the following substitution defined previously:
\begin{equation}
    \begin{split}
        k &= (l_{z} - a\epsilon)^{2} + q \\
        &= (l_{z} - a\epsilon)^{2} + l_{z}^{2} \frac{\Lambda^{2}}{1 - \Lambda^{2}} \, .
    \end{split}
\end{equation}

We remind the reader that $\Lambda \in [0, 1]$ is a parameter that measures the effective inclination (polar angle) of the particle's orbit at spatial infinity. Its limited range makes parametrisation of subsequent equations more convenient.

IBSOs satisfy $R(r = \chi) = R'(r = \chi) = 0$ and $\epsilon = 1$. These equations give a condition on $l_{z}$:
\begin{equation}
    l_{z} = \frac{1}{a(\chi - 1)} \left[ \chi (\chi - Q^{2}) - a^{2} - \chi^{1/2} (\chi^{2} - 2\chi + a^{2} + Q^{2}) \right] \, ,
\end{equation}
and a characteristic equation for $\chi$:
\begin{equation} \label{eq:ibso_kn}
    \begin{split}
        0 &= 3\chi^{6} - (Q^{2} + 4) \chi^{5} + a^{2}(9\Lambda^{2} - 2) \chi^{4} \\
        &\quad+ a^{2} \left( 4 - 16\Lambda^{2} + 4Q^{2} - 6Q^{2}\Lambda^{2} + 3Q^{4}/a^{2} \right) \chi^{3} \\
        &\quad+ a^{2} \left( -a^{2} + 6a^{2} \Lambda^{2} \right) \chi^{2} \\
        &\quad- a^{2} Q^{2} \left( 4 + 16\Lambda^{2} - Q^{2} + Q^{2} \Lambda^{2} - Q^{4} \right) \chi^{2} \\
        &\quad+ a^{2} \left( a^{2} Q^{2} - 2a^{2} \Lambda^{2} Q^{2} + Q^{4} - 4Q^{4} \Lambda^{2} \right) \chi \\
        &\quad+ a^{6} \Lambda^{2} \\
        &\quad- 2 (2\chi - Q^{2})^{2} \chi^{3/2} (\chi^{2} - 2\chi + a^{2} + Q^{2}) \, .
    \end{split}
\end{equation}

Given parameters $a$, $Q$, and $\Lambda$, it is straightforward to numerically solve the above equations for $\chi$ and a corresponding $l_{z}$. Solutions generally exist for both positive and negative $l_{z}$, corresponding to prograde and retrograde orbits respectively. For the same set of parameters and orbit orientation, multiple solutions of $\chi$ may exist. The largest one is taken as the IBSO radius.

\begin{figure*}[ht!]
    \centering
    \begin{subfigure}[b]{0.3275\textwidth}
        \centering
        \includegraphics[height=0.935\imageheight]{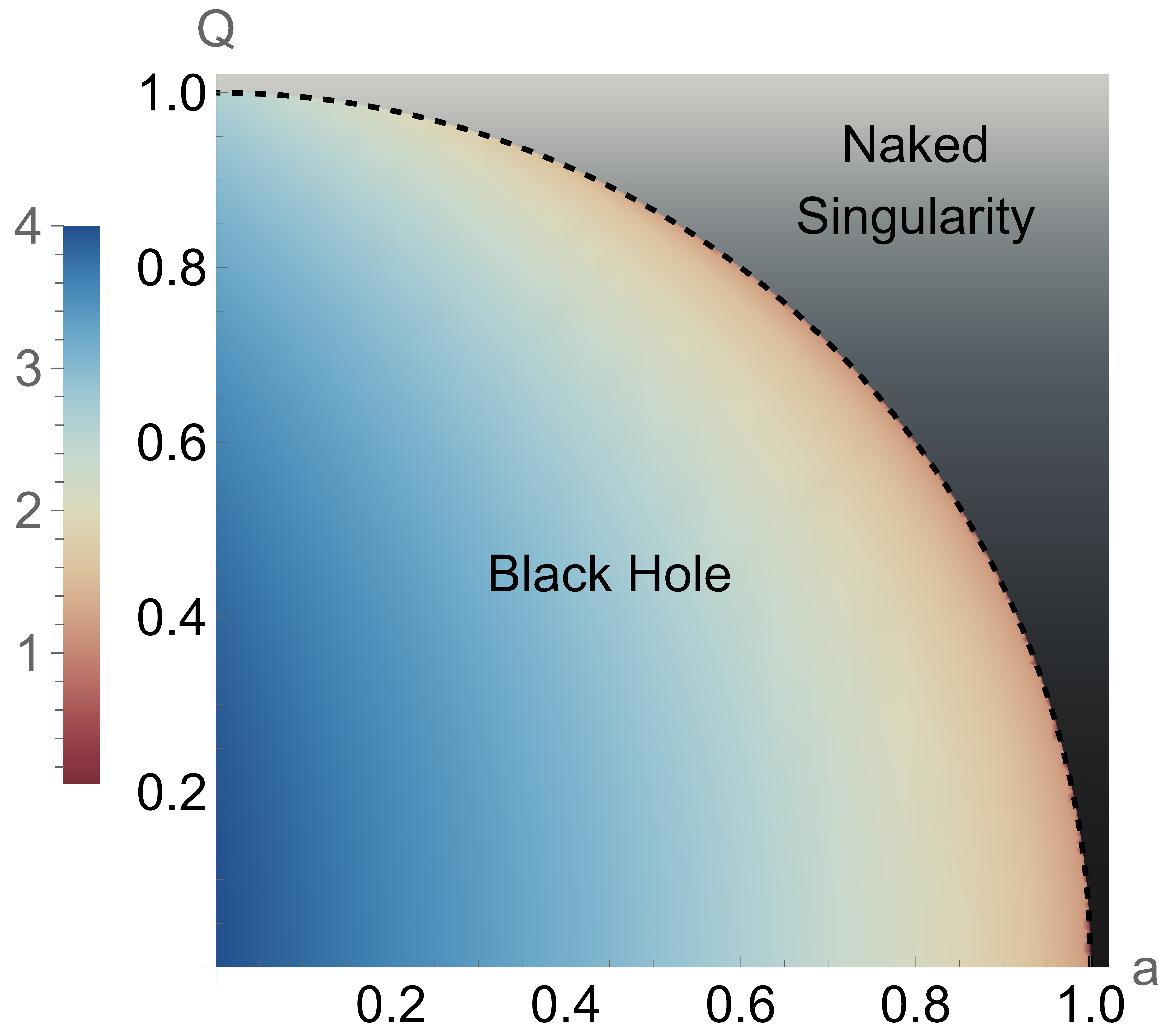}
        \caption{$\Lambda = 0.0$, prograde}
    \end{subfigure}
    \begin{subfigure}[b]{0.3275\textwidth}
        \centering
        \includegraphics[height=0.935\imageheight]{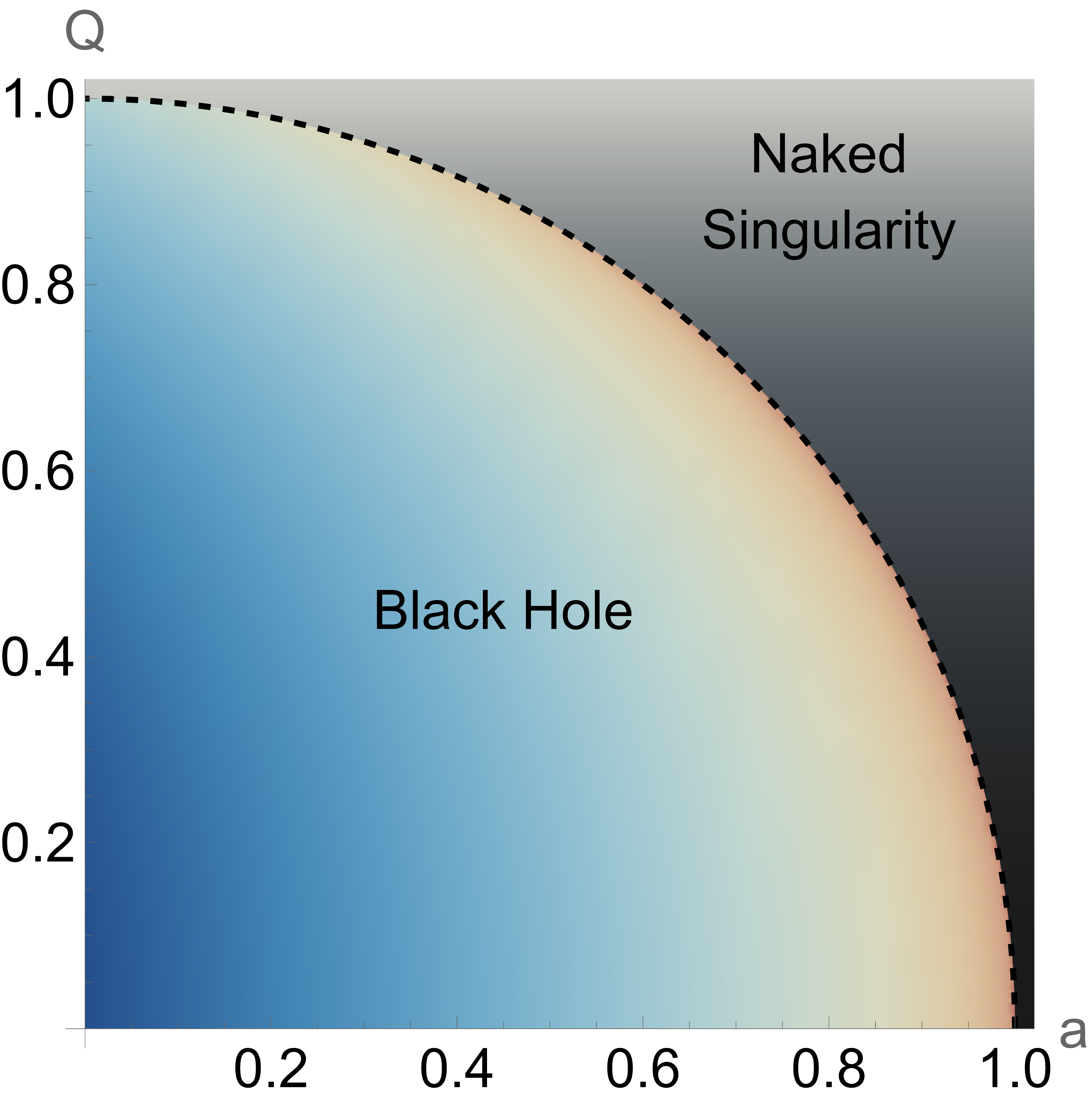}
        \caption{$\Lambda = 0.5$, prograde}
    \end{subfigure}
    \begin{subfigure}[b]{0.3275\textwidth}
        \centering
        \includegraphics[height=0.935\imageheight]{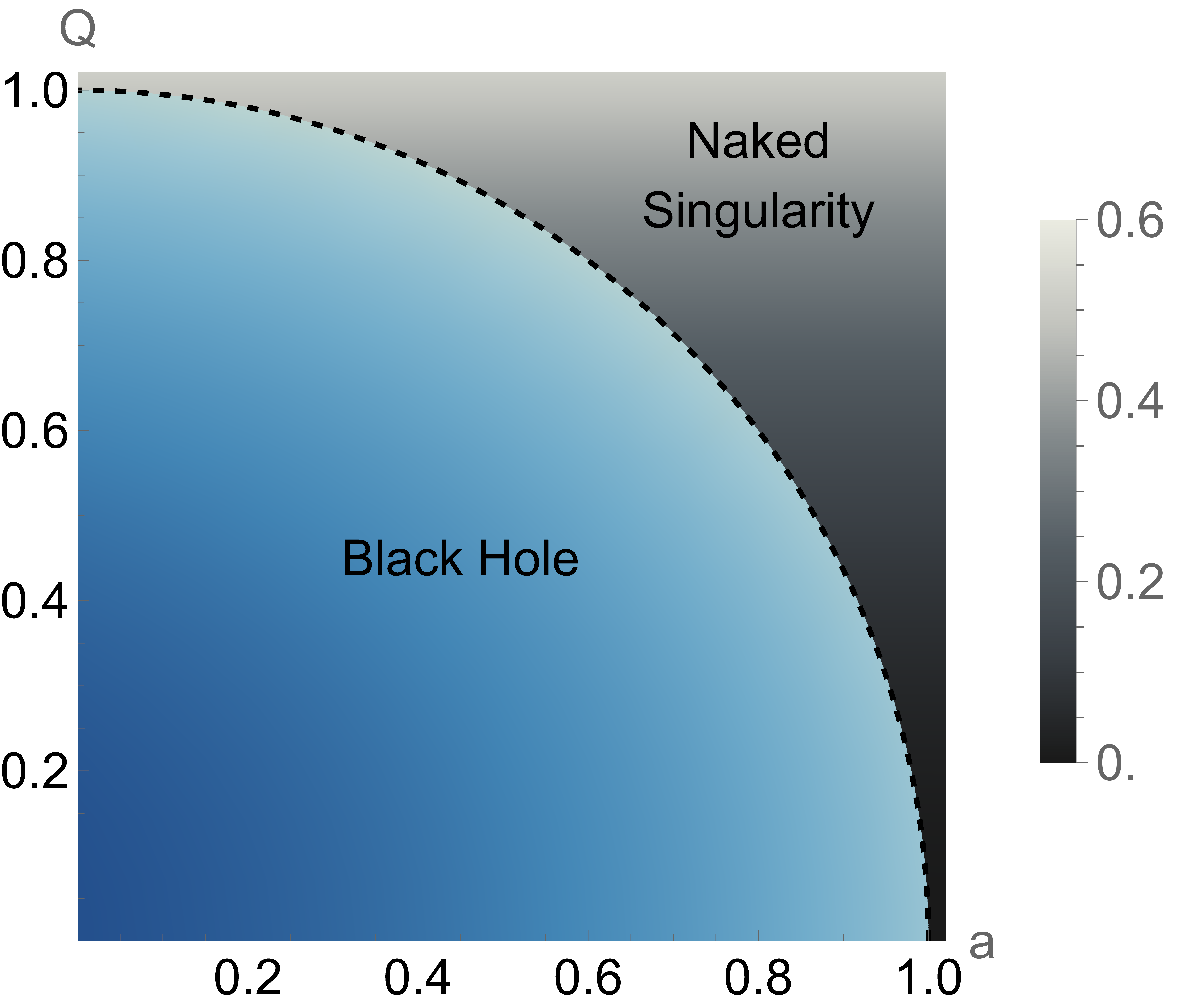}
        \caption{$\Lambda = 0.98$, prograde}
    \end{subfigure}

    \begin{subfigure}[b]{0.3275\textwidth}
        \centering
        \includegraphics[height=0.935\imageheight]{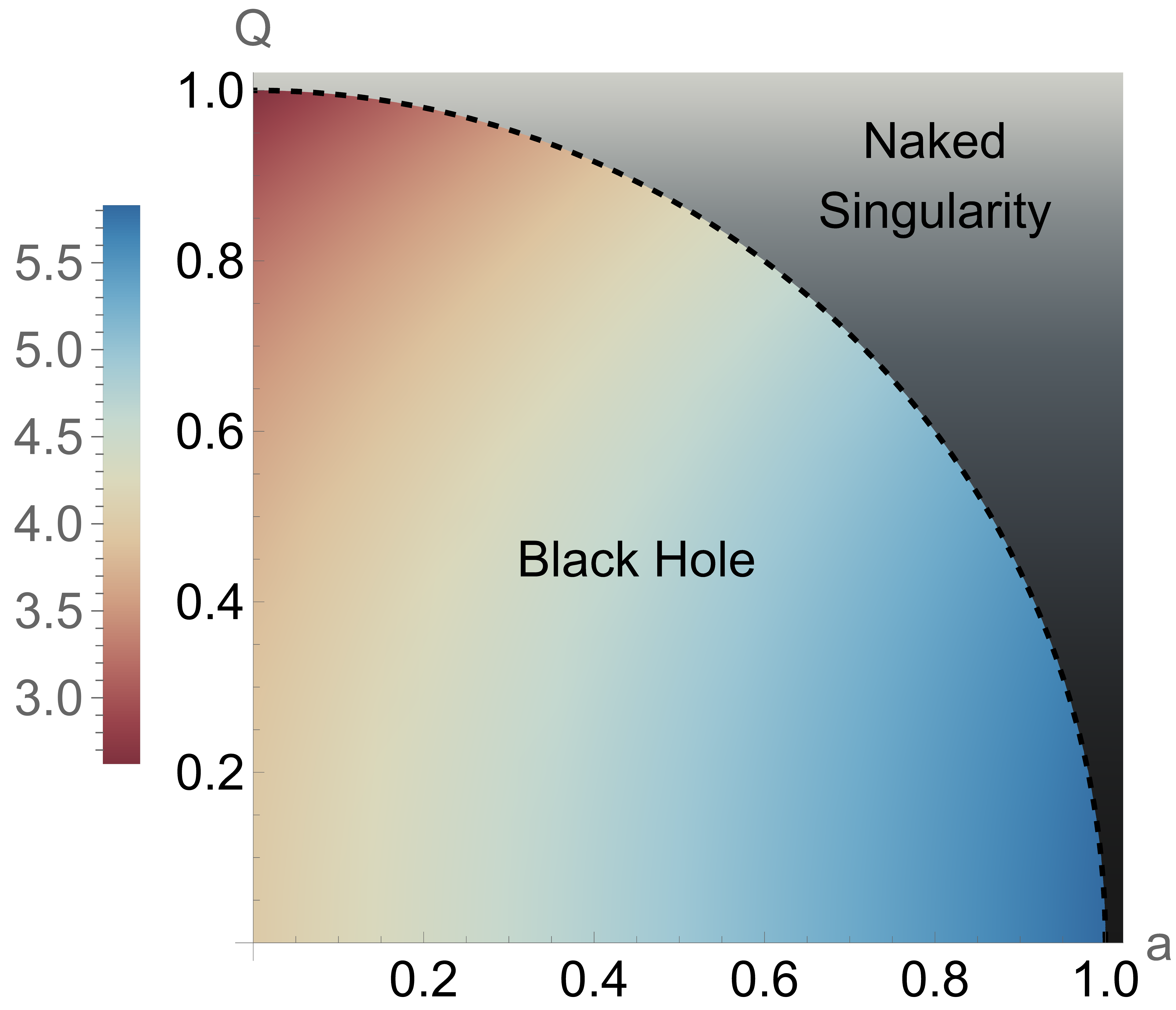}
        \caption{$\Lambda = 0.0$, retrograde}
    \end{subfigure}
    \begin{subfigure}[b]{0.3275\textwidth}
        \centering
        \includegraphics[height=0.935\imageheight]{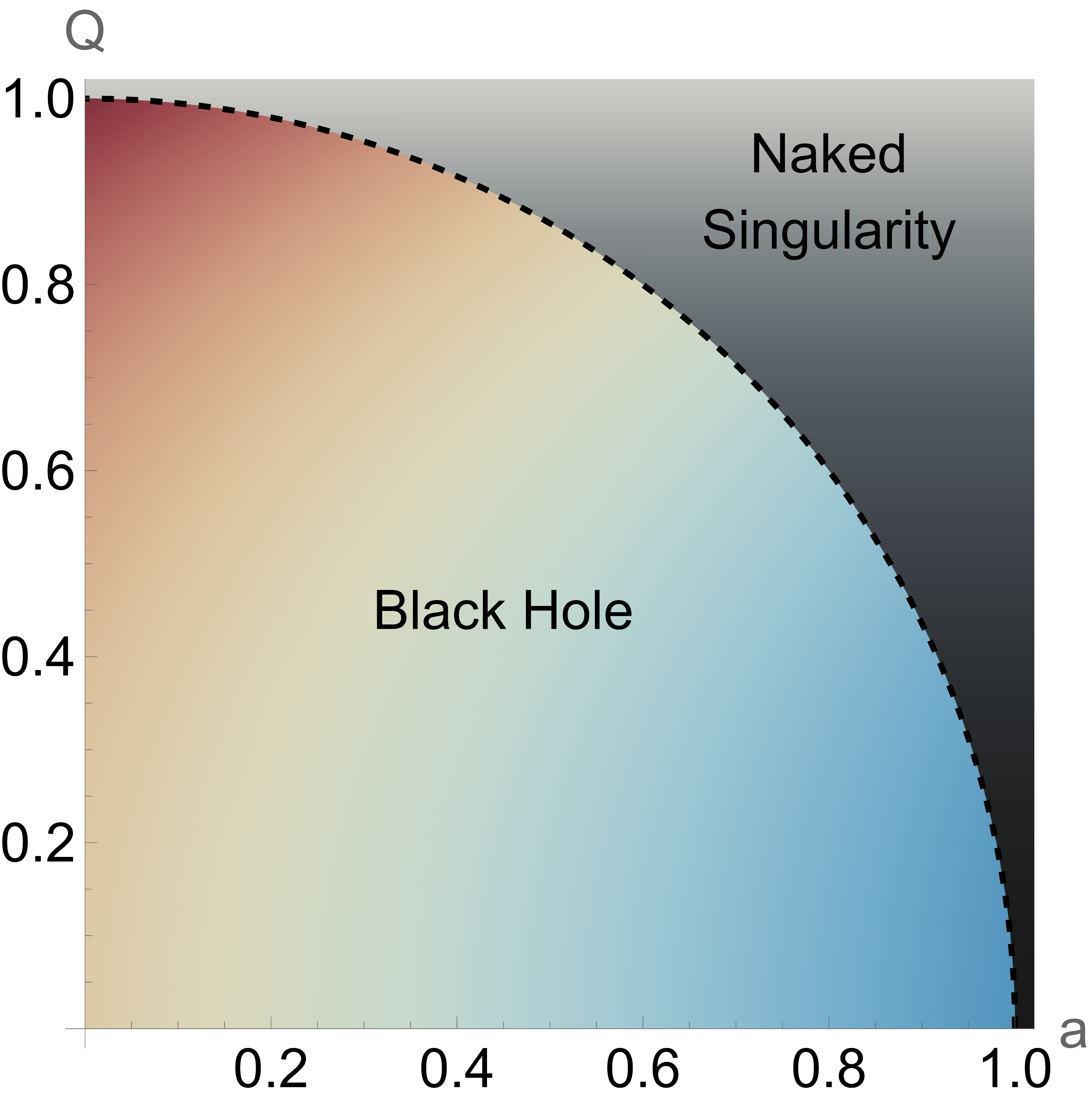}
        \caption{$\Lambda = 0.5$, retrograde}
    \end{subfigure}
    \begin{subfigure}[b]{0.3275\textwidth}
        \centering
        \includegraphics[height=0.935\imageheight]{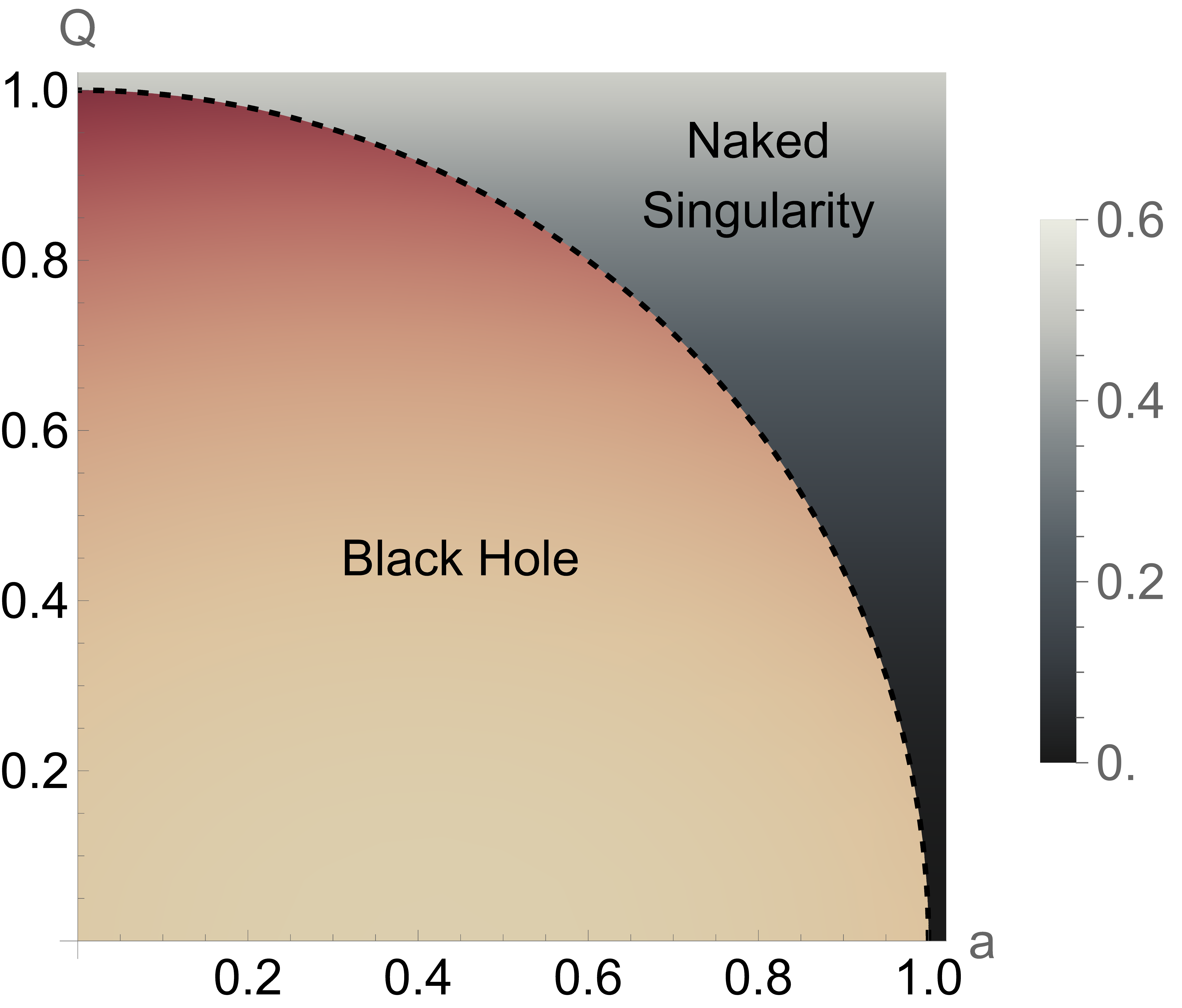}
        \caption{$\Lambda = 0.98$, retrograde}
    \end{subfigure}
    \caption{Heatmaps of radii of closest approach (IBSOs for BH; zero angular momentum orbits for NS) in Kerr-Newman spacetime. Black dashed line separates BH regime from NS regime ($a^2 + Q^2 = 1$). Different colour maps are used for the two regimes.}
    \label{fig:ibso_kn}
\end{figure*}

In the BH part of Figure \ref{fig:ibso_kn}, we plot the IBSO radii for prograde and retrograde orbits. In both cases, increasing $Q$ brings the orbit closer to the BH. Intuitively, this is because charge has a repulsive gravitational effect that allows a spherical orbit to be closer without plunging into the horizon. However, increasing $a$ brings prograde orbits closer but pushes retrograde orbits away. This is due to spin's frame dragging effect, which allows prograde orbits to get closer without falling into horizon, but has the opposite effect on retrograde orbits. Also note that the effect of charge is much less pronounced than the effect of spin.

\begin{figure*}[ht!]
    \centering
    \begin{subfigure}[b]{0.3275\textwidth}
        \centering
        \includegraphics[height=0.935\imageheightsecond]{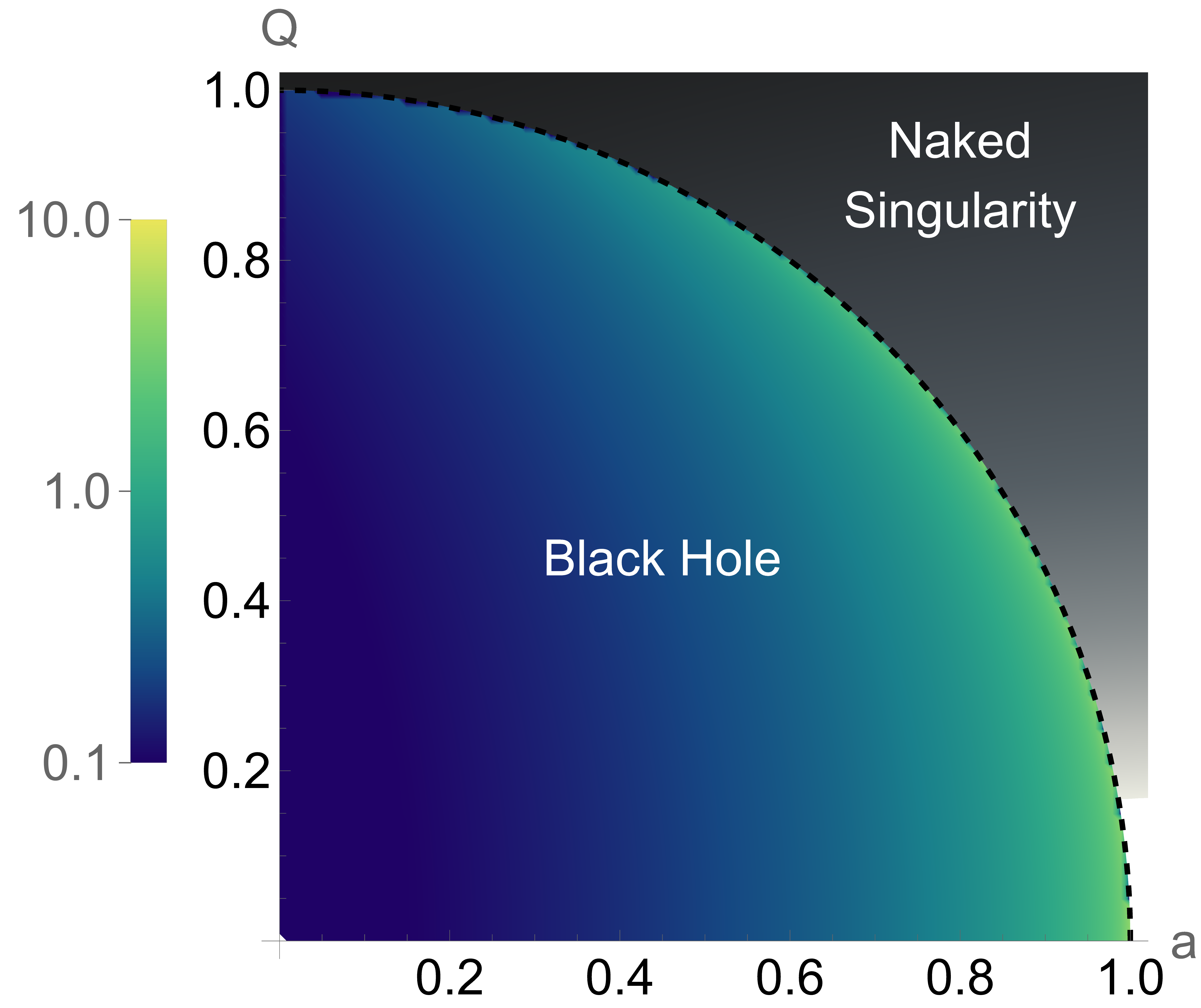}
        \caption{$\Lambda = 0.0$, prograde}
    \end{subfigure}
    \hfill
    \begin{subfigure}[b]{0.3275\textwidth}
        \centering
        \includegraphics[height=0.935\imageheightsecond]{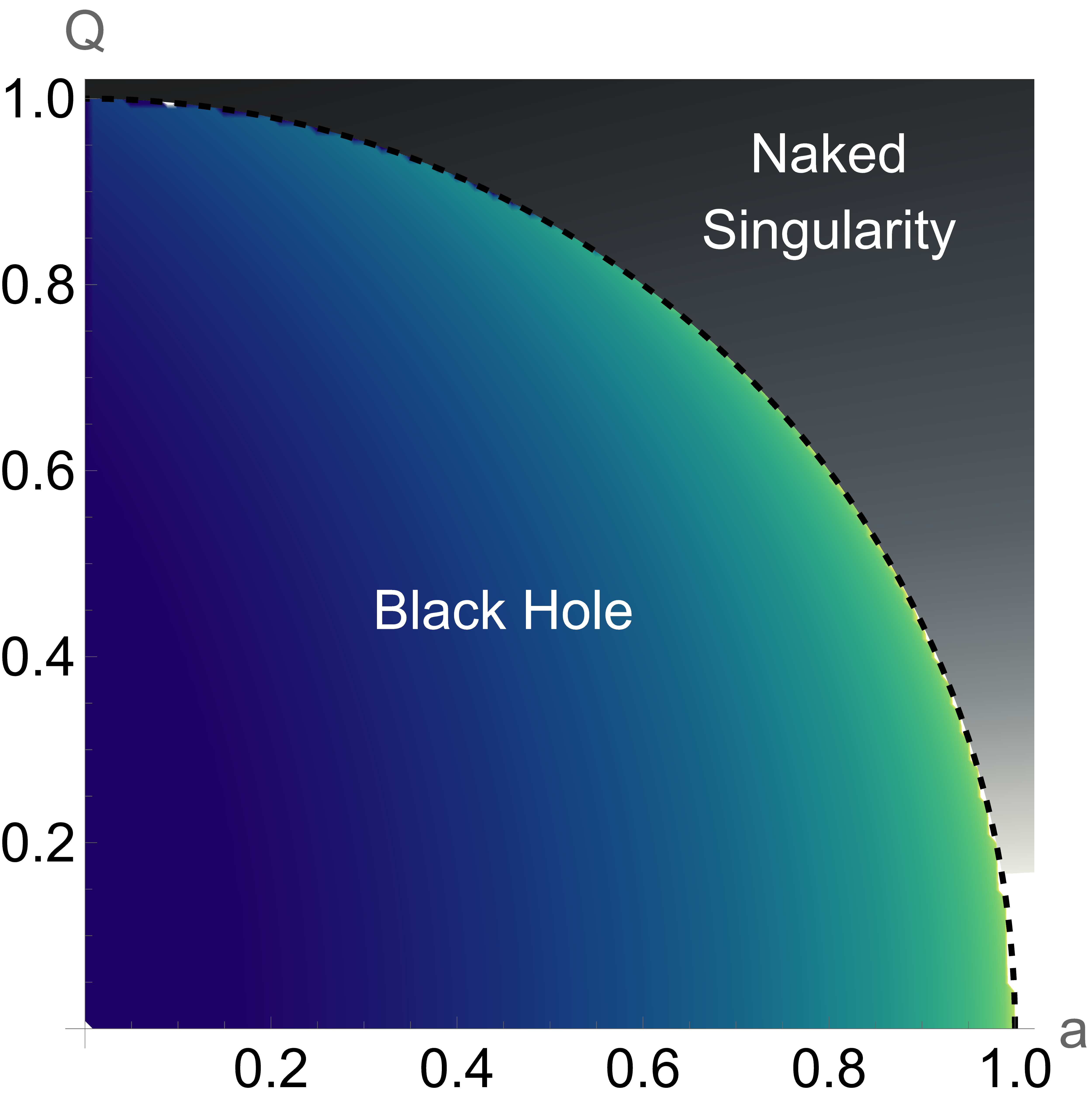}
        \caption{$\Lambda = 0.5$, prograde}
    \end{subfigure}
    \hfill
    \begin{subfigure}[b]{0.3275\textwidth}
        \centering
        \includegraphics[height=0.935\imageheightsecond]{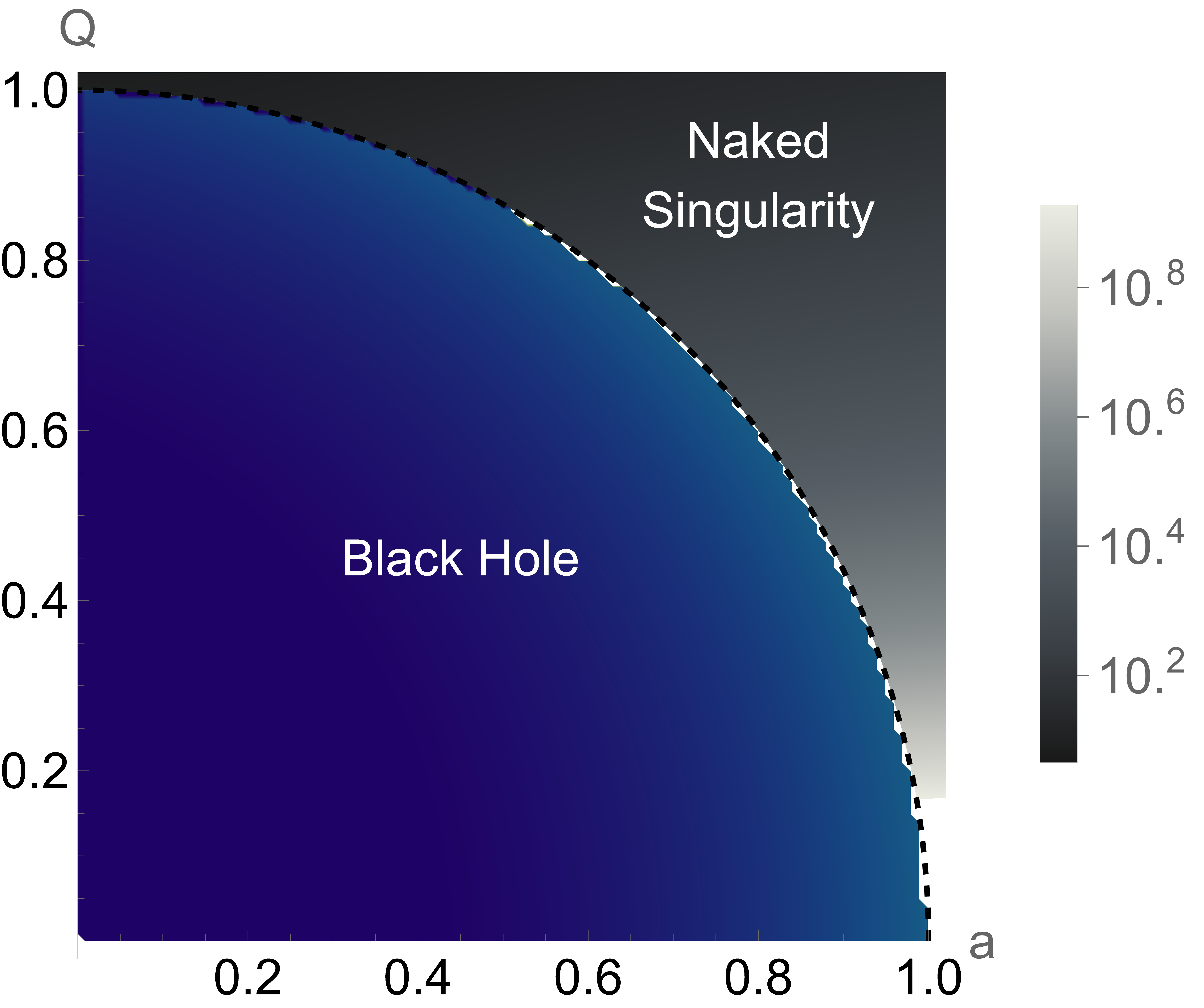}
        \caption{$\Lambda = 0.98$, prograde}
    \end{subfigure}

    \begin{subfigure}[b]{0.3275\textwidth}
        \centering
        \includegraphics[height=0.935\imageheightsecond]{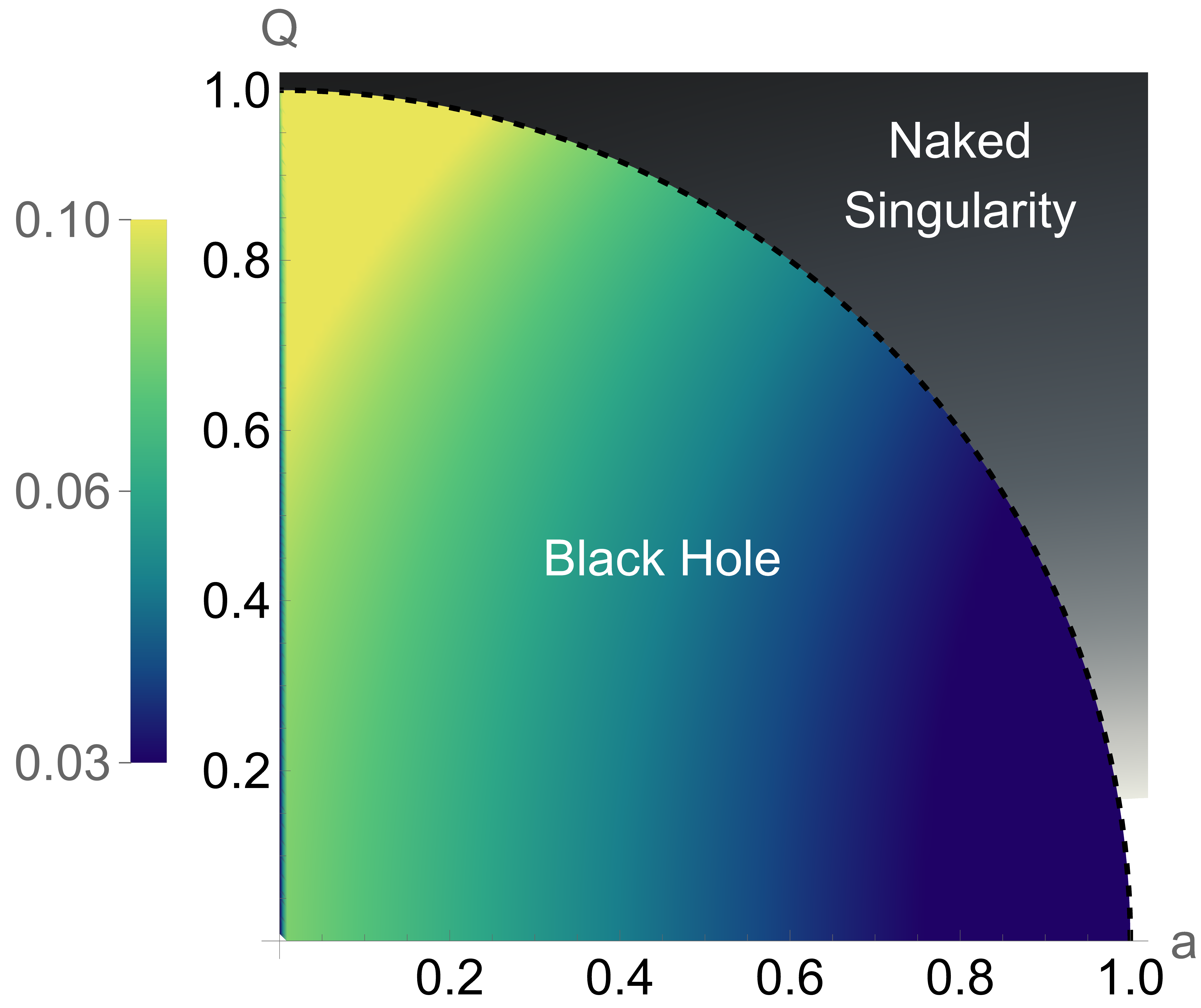}
        \caption{$\Lambda = 0.0$, retrograde}
    \end{subfigure}
    \hfill
    \begin{subfigure}[b]{0.3275\textwidth}
        \centering
        \includegraphics[height=0.935\imageheightsecond]{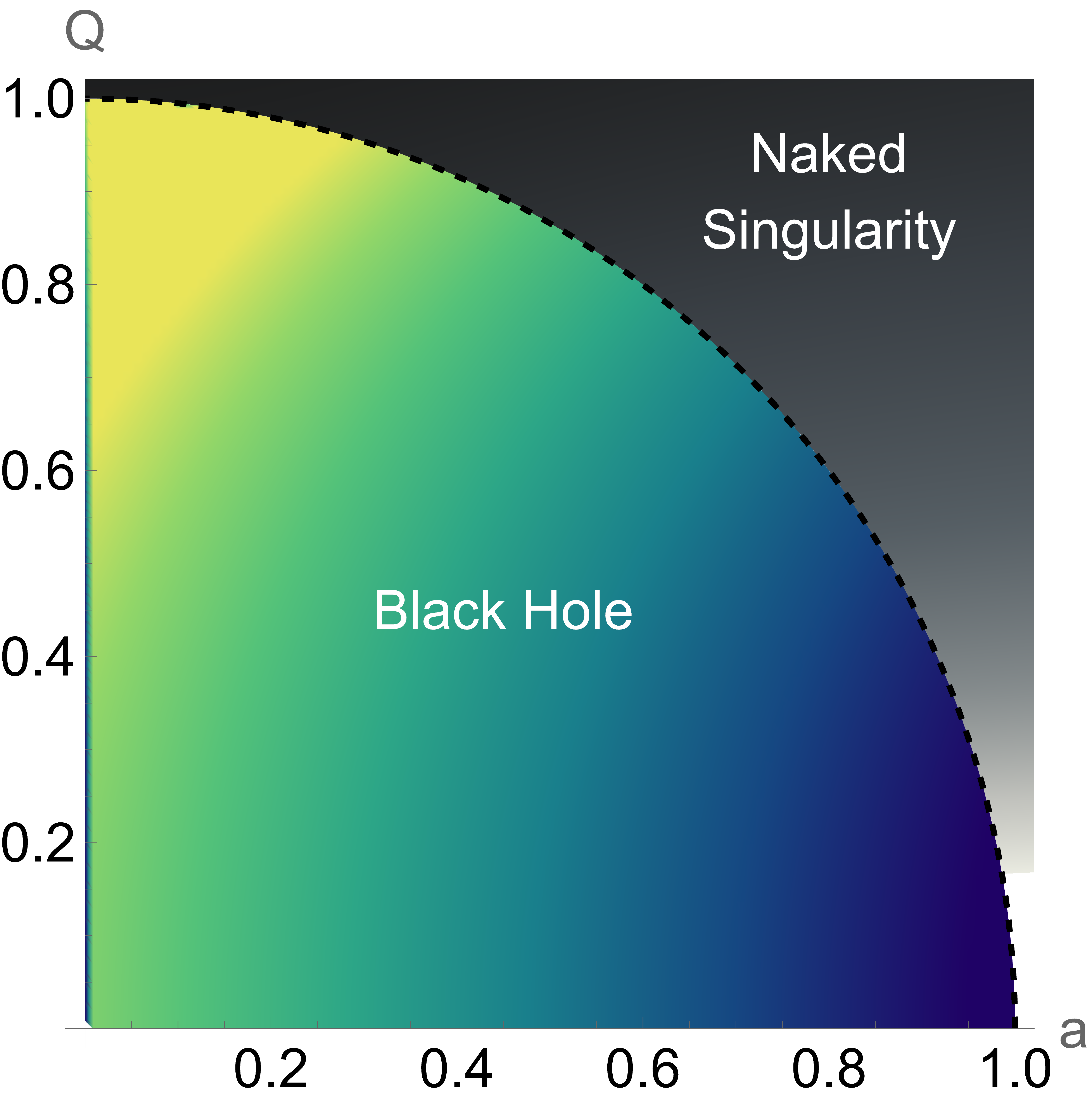}
        \caption{$\Lambda = 0.5$, retrograde}
    \end{subfigure}
    \hfill
    \begin{subfigure}[b]{0.3275\textwidth}
        \centering
        \includegraphics[height=0.935\imageheightsecond]{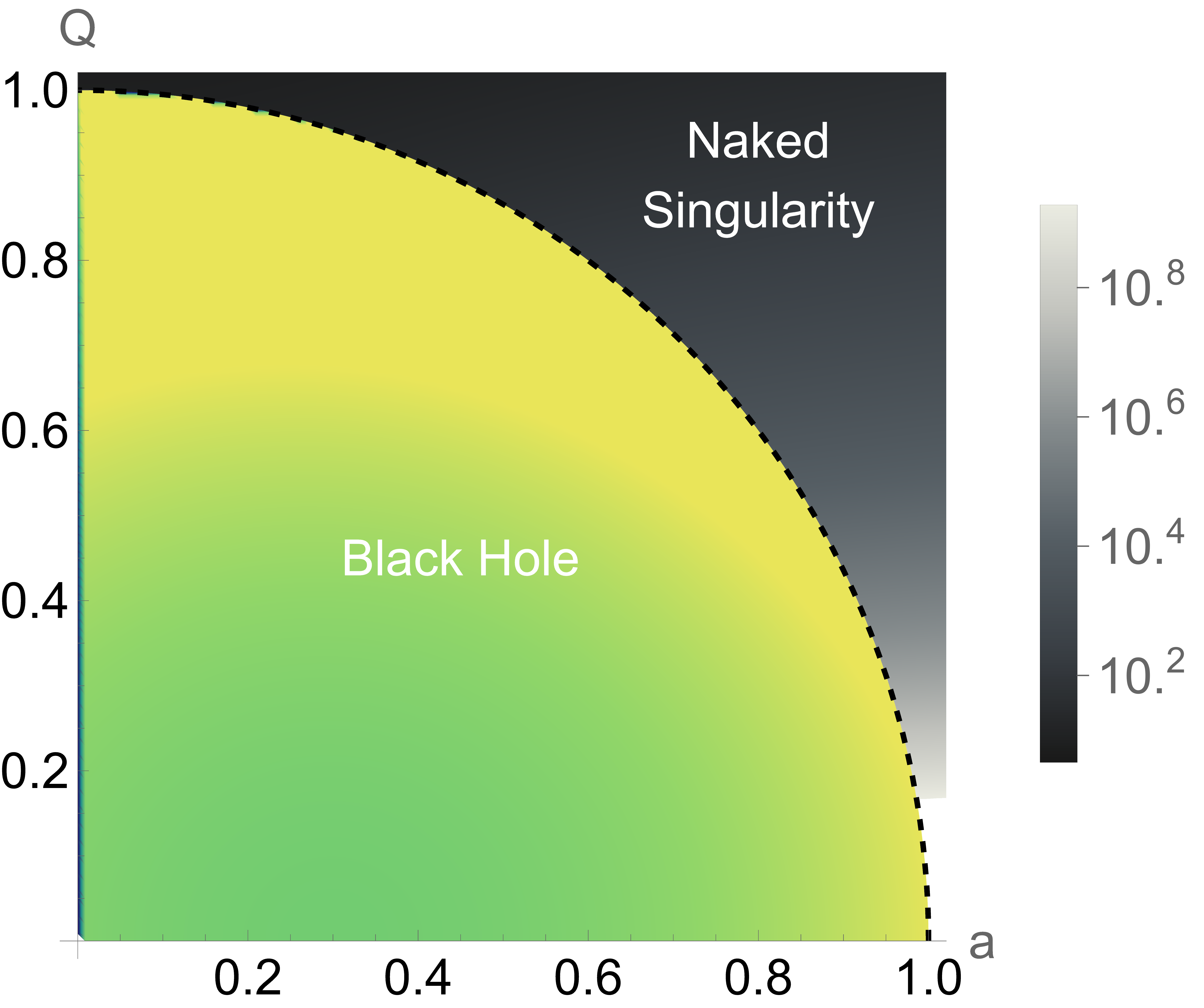}
        \caption{$\Lambda = 0.98$, retrograde}
    \end{subfigure}
    \caption{Heatmaps of the strongest stretching tidal eigenvalue in Kerr-Newman spacetime. Black dashed line separates BH regime from NS regime ($a^2 + Q^2 = 1$). Different colour maps are used for the two regimes. All eigenvalues are evaluated as the particle crosses the equatorial plane at the closest approach (IBSOs for BH; zero angular momentum orbits for NS).}
    \label{fig:ibso_eigenvalue_kn}
\end{figure*}

Since the IBSO radius is an implicit function of $a$, $Q$, and $\Lambda$, the tidal eigenvalues along the IBSOs (and thus the Hills mass) are also determined by these parameters. We numerically evaluate the tidal eigenvalues for particles on IBSO trajectories as they cross the equatorial plane ($\theta = \pi/2$) since this is where the strongest tidal forces occur. The strongest stretching (negative) tidal eigenvalues are plotted in the BH part of Figure \ref{fig:ibso_eigenvalue_kn}.

For a given KN BH specified by $a$ and $Q$, stars may approach from random inclination angles $\cos^{-1}{\Lambda}$ on either prograde or retrograde orbits. Therefore, the Hills mass must be defined by the strongest tidal eigenvalue across all possible $\Lambda$ on either prograde or retrograde orbits. Since tidal accelerations are generally stronger for prograde orbits, we can find the Hills mass of a KN BH by maximising the strongest stretching tidal eigenvalue on prograde IBSOs over $\Lambda \in [0, 1]$. We demonstrate this process below by considering extremal KN BHs. Note that, however, the procedure of maximising over $\Lambda$ is in general possible for any combination of $a$ and $Q$ in the BH regime.

\subsubsection{Extremal KN black holes}

An extremal KN BH satisfies $a^{2} + Q^{2} = 1$, which represents the boundary separating the BH regime from the NS regime. As extremal BHs have the smallest IBSO radii, their IBSOs produce the strongest tidal forces a particle can experience around a KN BH across the entire possible range of $a$ and $Q$.

\begin{figure}[ht]
    \centering
    \includegraphics[width=\columnwidth]{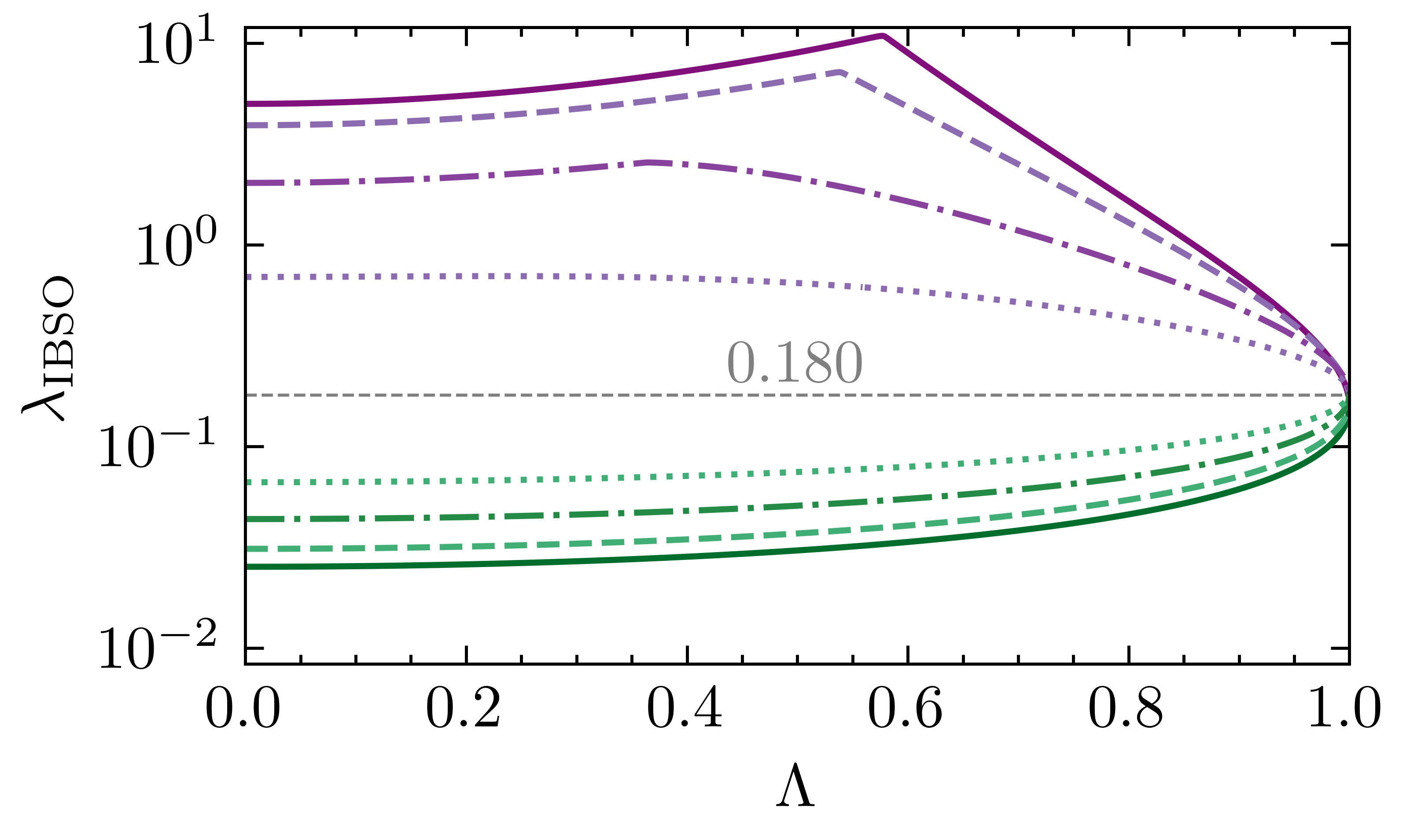}
    \caption{Magnitude of stretching tidal eigenvalues on IBSOs around extremal Kerr-Newman BHs as a function of inclination parameter $\Lambda$ for prograde orbits (purple) and retrograde orbits (green) for various values of charge parameter $Q = \sqrt{1 - a^{2}}$, with $Q = 0.0$ (solid), $Q = 0.5$ (dashed), $Q = 0.75$ (dash-dotted), and $Q = 0.9$ (dotted). All eigenvalues are evaluated as the particle crosses the equatorial plane.}
    \label{fig:maximal_kn}
\end{figure}

Figure \ref{fig:maximal_kn} shows the magnitude of stretching (negative) tidal eigenvalues on IBSOs around extremal KN BHs plotted against inclination parameter $\Lambda$ for various values of $Q$. Note that tidal eigenvalues on prograde orbits always have larger magnitudes than those on retrograde orbits. Thus, the retrograde curves are irrelevant for determining the critical Hills mass of an extremal KN BH.

All of the prograde and retrograde curves converge to the same eigenvalue at $\Lambda = 1$, which corresponds to polar orbits, regardless of the value of $Q$. This eigenvalue is identical to the one calculated for an IBCO around an extremal RN BH (Section \ref{sec:spherical}).

We observe that for prograde curves where $Q$ is not too large, there are discontinuities at certain values of $\Lambda$ where the magnitude of the eigenvalue is maximised. This implies that stars are most susceptible to tidal disruption when approaching a KN BH at specific inclination angles, not necessarily on equatorial orbits. The discontinuity determines the Hills mass for an extremal KN BH of a given $Q = \sqrt{1 - a^{2}}$. If the value of $Q$ is large enough, the discontinuity disappears and the eigenvalue is maximised at $\Lambda = 0$, i.e., equatorial orbits.

The discontinuities, or ``kinks'', in the tidal eigenvalues and the Hills mass have been noted in the context of Kerr BHs \cite{Mummery2023}. For a Kerr BH with an extremal spin parameter, there is a ``phase transition'' to the solution of the Kerr IBSO characteristic equation at some critical inclination angle \cite{Hod2013}, which manifests as a discontinuity in the Hills mass curve. Figure \ref{fig:maximal_kn} generalises this observation to extremal KN BHs.

Following \cite{Mummery2023}, the critical inclination angle for an extremal Kerr BH is found to be $\Lambda_{\text{crit}} = 1/\sqrt{3} \approx 0.577$ and the corresponding stretching (negative) eigenvalue is $-11$, both of which agree with Figure \ref{fig:maximal_kn} when $Q = 0$.

\begin{figure}[ht]
    \centering
    \includegraphics[width=\columnwidth]{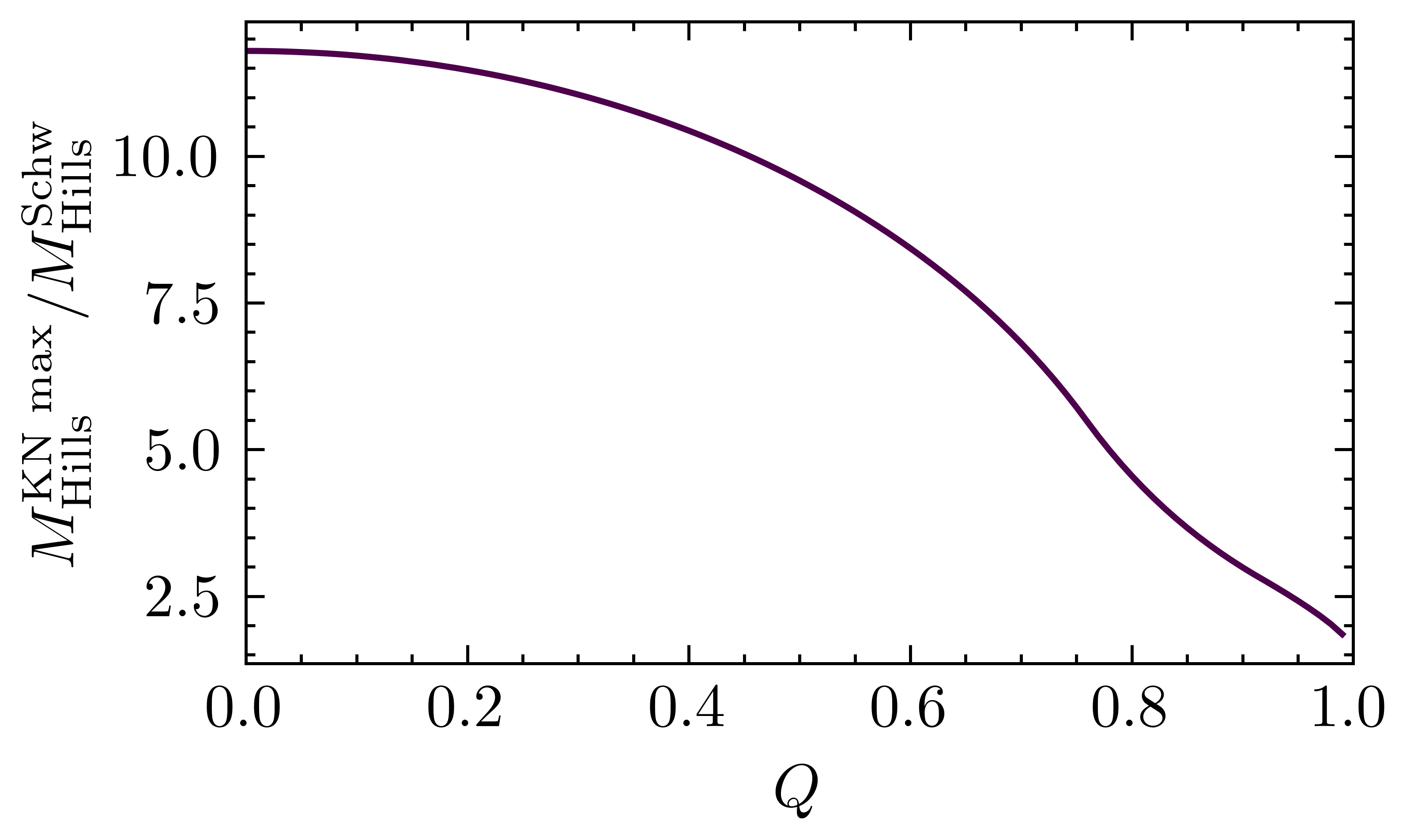}
    \caption{Hills mass of an extremal KN BH (in units of the Schwarzschild Hills mass) as a function of the charge parameter Q.}
    \label{fig:kn_hills}
\end{figure}

We further extract the maxima of the prograde curves in Figure \ref{fig:maximal_kn} to plot the Hills mass of an extremal KN BH as a function of $Q$ in Figure \ref{fig:kn_hills}. The curve has been scaled using the Schwarzschild Hills mass $M_{\text{Hills}}^{\text{Schw}} = (\sqrt{5}/8) M_{0}$. The $Q = 0$ end represents the extremal Kerr BH case with a Hills mass of:
\begin{equation}
    M_{\text{Hills}}^{\text{Kerr max}} = \sqrt{11} M_{0} \, ,
\end{equation}
which is about $11.8$ times the Schwarzschild Hills mass, as reported in \cite{Mummery2023}.

The $Q = 1$ end represents the extremal RN BH case, which we have shown in Section \ref{sec:spherical} to have a Hills mass of:
\begin{equation}
    M_{\text{Hills}}^{\text{RN max}} = (5\sqrt{5} - 11)^{1/2} M_{0} \, ,
\end{equation}
which is about $1.52$ times the Schwarzschild Hills mass.

\subsubsection{KN naked singularities}

We now elaborate on the NS region of Figures \ref{fig:ibso_kn} and \ref{fig:ibso_eigenvalue_kn}. When $Q^{2} + a^{2} > 1$, the metric describes a NS without a horizon. As explained before, we set $\epsilon = 1$ and $l_{z} = 0$ and notice that the radial potential can be factored:
\begin{equation}
    R(r) = 2 r^{3} - Q^{2} r^{2} + 2 a^{2} r - a^{2} Q^{2} = (2r - Q^{2}) (r^{2} + a^{2}) \, .
\end{equation}
This implies that there is always a single positive root at $r = Q^{2}/2$, independent of $a$. Perhaps unexpectedly, this is exactly the same as in the RN spacetime.

Unlike Kerr, a NS in KN spacetime can never be reached by a particle with unit energy. There hence exists an upper limit on the tidal forces that can be experienced by a particle around a KN NS. This implies a finite Hills mass.

It is possible to derive the analytical expression of the local tidal tensor at this closest approach for a particle travelling in a head-on collision course on the equatorial plane. Setting $r = Q^{2}/2$, $\theta = \pi/2$, $\epsilon = 1$, $l_{z} = 0$, and $k = a^{2}$, the tidal eigenvalues are:
\begin{equation}
    \begin{split}
        \lambda_{1} &= \frac{8}{Q^{10}} (20a^{2} + 4Q^{4}) \, , \\
        \lambda_{2} &= -\frac{8}{Q^{10}} (4a^{2} + Q^{4}) \, , \\
        \lambda_{3} &= -\frac{8}{Q^{6}} \, .
    \end{split}
\end{equation}
Note how the results reduce to the NS case in the RN spacetime (Equations \ref{eq:rn_eigenvalues}) when $a = 0$: one compressing eigenvalue of the magnitude $32/Q^{6}$ and two stretching ones of $-8/Q^{6}$. Also note that the eigenvalues diverge as $Q \to 0$, which is expected as in this limit the spacetime reduces to a Kerr NS, which has an infinite Hills mass.

Using the strongest stretching eigenvalue $\lambda_{2}$, we can compute the Hills mass for a KN NS:
\begin{equation}
\begin{split}
    M_{\text{Hills}}^{\text{KN NS}} &= \left[ \frac{8}{Q^{10}} (4a^{2} + Q^{4}) \right]^{1/2} M_{0} \\
    &= \left( 1 + \frac{4 a^{2}}{Q^{4}} \right)^{1/2} M_{\text{Hills}}^{\text{RN NS}} \, ,
\end{split}
\end{equation}
where we have used the RN NS Hills mass from Eq. \ref{eq:rn_ns_hills_mass}.

\section{Exotic Spacetime --- Rotating Wormhole} \label{sec:wormhole}

As a final example to illustrate the versatility of the ZAMO method, let us investigate a special instance of wormhole spacetimes. Teo \cite{Teo1998} showed that a stationary, axisymmetric wormhole can be described by the metric functions (see Equation \ref{eq:metric}):
\begin{equation} \label{eq:wormhole_metric}
    \begin{split}
        &e^{2\nu} = [N(r, \theta)]^{2} \, , \qquad e^{2\psi} = r^{2} \sin^{2}{\theta} [K(r, \theta)]^{2} \, , \\
        &e^{2\mu_{1}} = \left[ 1 - \frac{b(r, \theta)}{r} \right]^{-1} \, , \qquad e^{2\mu_{2}} = r^{2} [K(r, \theta)]^{2} \, , \\
        &\omega = \omega(r, \theta) \, ,
    \end{split}
\end{equation}
where $N$, $K$, $b$, and $\omega$ are prescribed functions of $r$ and $\theta$.

For simplicity, consider the particular case proposed by Teo:
\begin{equation}
    N = K = 1 + \frac{(4a\cos{\theta})^{2}}{r} \, , \qquad b = 1 \, , \qquad \omega = \frac{2a}{r^{3}} \, ,
\end{equation}
where the parameter $a$ is a measure of the wormhole's angular momentum.

We consider a particle on the equatorial plane $\theta = \pi/2$ as this allows for analytical expressions of the 4-velocities. Since the particle has energy and angular momentum as conserved quantities, the first integrals can be written down \cite{Teo1998}:
\begin{equation}
    \begin{split}
        \dot{t} &= \frac{1}{D^{2}} (g_{\phi\phi} \epsilon + g_{t\phi} l_{z}) \, , \\
        \dot{r}^{2} &= g^{rr} \left[ \frac{1}{D^{2}} (g_{\phi\phi} \epsilon^{2} + 2g_{t\phi} \epsilon l_{z} + g_{tt} l_{z}^{2}) - 1 \right] \, , \\
        \dot{\theta} &= 0 \, , \\
        \dot{\phi} &= -\frac{1}{D^{2}} (g_{t\phi} \epsilon + g_{tt} l_{z}) \, ,
    \end{split}
\end{equation}
where the equation for $\dot{r}$ results from normalising the 4-momenta and $D$ is defined as:
\begin{equation}
    D \equiv \sqrt{-g_{tt} g_{\phi\phi} + g_{t\phi}^{2}} = NKr\sin{\theta} \, .
\end{equation}
Given the first integrals, it is possible to derive a local tidal tensor parametrised by $\epsilon$ and $l_{z}$. However, the expressions are too convoluted to be insightful so we instead set $\epsilon = 1$ and $l_{z} = 0$. This represents a particle (spacecraft) from spatial infinity on a head-on collision course, whose local tidal tensor takes a particularly simple form:
\begin{equation}
    C =
    \frac{a^{2}}{r^{7}}
    \begin{bmatrix}
        0 & 0         & 0        & 0         \\
        0 & 27(r - 1) & 0        & 0         \\
        0 & 0         & -32r^{4} & 0         \\
        0 & 0         & 0        & -9(r - 1)
    \end{bmatrix} \, .
\end{equation}
Note that tidal accelerations vanish if $a = 0$. At the ``throat'' of the wormhole $r = 1$, the only non-vanishing eigenvalue is $\tensor{C}{^{2}_{2}} = -32a^{2}$ pointing at the polar direction.

Setting $\epsilon \neq 1$ leads to non-vanishing tidal accelerations even if $a = 0$, in which case the local tidal tensor becomes:
\begin{equation}
    C =
    \frac{\epsilon^{2} - 1}{2r^{3}}
    \begin{bmatrix}
        0 & 0 & 0 & 0 \\
        0 & 0 & 0 & 0 \\
        0 & 0 & 1 & 0 \\
        0 & 0 & 0 & 1
    \end{bmatrix} \, ,
\end{equation}
i.e., an interstellar traveller has to be mindful of how fast they approach a wormhole.

The interpretation of the local tidal tensor \textit{vis-à-vis} the traversability of a wormhole is outside the scope of the present study. However, the ZAMO method could serve as a useful tool to verify the ``engineering feasibility'' of a wormhole spacetime (see \cite{Lobo2007} for a review).

\section{Conclusions} \label{sec:conclusion}

We have constructed a general analytical framework for deriving the local tidal tensor in a stationary, axisymmetric spacetime using the ZAMO tetrad and standard relativistic frame transformations. To the best of our knowledge, this is the first attempt at deriving a general procedure for computing the relativistic tidal tensor. We have shown that the method yields a completely general analytical local tidal tensor in spherical spacetimes; it can also be coupled with simple numerical geodesic integration methods to simulate tidal accelerations in more complicated cases such as the Kerr-Newman spacetime.

As an example of the astrophysical implication of our work we have computed the ``Hills mass'' (the maximum mass of a compact object which can tidally disrupt a star and produce observable emission) in a number of extended spacetimes. We find a number of interesting, and potentially surprising, results: 

\begin{itemize}
    \item Despite representing naked singularity spacetimes, any super-extremal Kerr-Newman spacetime $(a^2 + Q^2 > 1)$ has a finite Hills mass, with value of $M_{\text{Hills}}^{\text{KN NS}} = (1+4a^{2}/Q^{4})^{1/2} (8c^{6} R_{\star}^{3}/\eta G^{3}M_{\star}Q^{6})^{1/2}$.
    \item Kerr-like naked singularities $(a^2>1)$ however have formally infinite Hills masses, as a unit-energy test particle can get arbitrarily close to $r=0$ on a zero angular momentum orbit. The observed super-exponential cut-off in detected TDE emission above $M \sim 10^8 M_\odot$ seen in TDEs \cite{Yao23, MummeryVV25} can likely be used to place tight constraints on the prevalence of such objects in the Universe. 
\end{itemize}

We have, in this work, only considered the possibility of a TDE occurring (i.e., if detectable emission could in principle be detected from the disruption) but not the properties of the disruption process itself. A natural extension of this work would be to consider the eigenvectors of the tidal forces in these extended metrics, such that the subsequent evolution of the tidally stripped debris streams can be studied. In a more mundane setting, the rapid computation of tidal accelerations magnitudes and directions allowed by our ZAMO approach may aid in speeding up relativistic tidal force calculations in the Kerr spacetime. 


\begin{acknowledgments}
    We would like to thank Garret Cotter for his illuminating discussion and suggestions. We wish to thank the referee for a positive report, and suggestions which improved the manuscript. 
\end{acknowledgments}


\appendix
\section{Local tidal tensor in spherical spacetime}\label{sec:app}

Assuming a general spherical metric described by Equation \ref{eq:spherical_metric} and setting $\theta = \pi/2$ and $\epsilon = 1$, the local tidal tensor can be written as:
\begin{equation}
    C =
    \begin{bmatrix}
        0 & 0         & 0         & 0         \\
        0 & \tensor{C}{^{1}_{1}} & 0         & \tensor{C}{^{1}_{3}} \\
        0 & 0         & \tensor{C}{^{2}_{2}} & 0         \\
        0 & \tensor{C}{^{3}_{1}} & 0         & \tensor{C}{^{3}_{3}}
    \end{bmatrix} \, ,
    \label{eq:tidal_tensor_spherical}
\end{equation}
where:
\begin{equation}
    \begin{split}
        f \tensor{C}{^{1}_{1}} &= 2 r^{3} A^{1/2} B \left( l_{z}^{2} + r^{2} \right) A'^{2} + r^5 B A'^{2} \\
        &\quad+ A^{2} r B' \left( \left( l_{z}^{2} + r^{2} \right)^{2} A'-2 l_{z}^{2} r \right) \\
        &\quad+ 2A^{2} B \left( l_{z}^{2} + r^{2} \right) \left( l_{z}^{2} A'-r \left( l_{z}^{2} + r^{2} \right) A'' \right) \\
        &\quad+ r A \left( A' \left( B \left( \left( l_{z}^{2} + r^{2} \right)^{2} A'-2 l_{z}^{2} r \right) + r^{4} B' \right) \right) \\
        &\quad-2 r^{5} A B A'' \\
        &\quad+ 2 r^{3} A^{3/2} \left( l_{z}^{2} + r^{2} \right) \left( A' B'-2 B A'' \right) \\
        &\quad-4 l_{z}^{2} r^{2} A^{5/2} B'-2 l_{z}^{2} r^{2} A^{3} B' \, ,
    \end{split}
\end{equation}
\begin{equation}
    \begin{split}
        (2 r^{4} A^{2} B^{2}) \tensor{C}{^{2}_{2}} &= B \left( r^3 \left( -A' \right)-2 l_{z}^{2} A^{2} (B-1) \right) \\
        &\quad+r A \left( A \left( l_{z}^{2}+r^{2} \right)-r^{2} \right) B' \, ,
    \end{split}
\end{equation}
\begin{equation}
    \begin{split}
        f \tensor{C}{^{3}_{3}} &= r A \left( A \left( l_{z}^{2}+r^{2} \right)-r^{2}\right) B' \left( 2 r ( A^{1/2} +1 )^{2}-l_{z}^{2} A'\right) \\
        &\quad+B \left( 2 l_{z}^{2} r A \left( A \left( l_{z}^{2}+r^{2} \right)-r^{2} \right) A'' \right) \\
        &\quad+B \left( l_{z}^{2} r \left( r^{2}-A \left( l_{z}^{2}+r^{2} \right) \right) A'^{2} \right) \\
        &\quad-2BA' \left( -l_{z}^{2} A+r^{2} A^{1/2} +r^{2} \right)^{2} \, ,
    \end{split}
\end{equation}
\begin{equation}
    \begin{split}
        &f \frac{A^{1/2}B^{1/2}}{l_{z} r} \left( \frac{B}{1/A - l_{z}^{2}/r^{2} - 1} \right)^{1/2} \tensor{C}{^{1}_{3}} = \\
        &- r A^{1/2} B A' \left( \left( l_{z}^{2} + r^{2} \right) A' - 2 r \right) \\
        &+ A^{3/2} \left( 2 B \left( r \left( l_{z}^{2} + r^{2} \right) A'' - l_{z}^{2} A' \right) \right) \\
        &- A^{3/2} r B' \left( \left( \left( l_{z}^{2} + r^{2} \right) A' - 2 r \right) \right)  \\
        &+ r^{2} A \left( 2 B \left( r A'' + A' \right) - r A' B' \right) + 2 r^{2} A^{5/2} B' \\
        &+ 4 r^{2} A^{2} B' - r^3 B A'^{2} \, ,
    \end{split}
\end{equation}
\begin{equation}
    \tensor{C}{^{3}_{1}} = \tensor{C}{^{1}_{3}} \, ,
\end{equation}
and:
\begin{equation}
    f(A, B) = 4 r^5 \left( A^{1/2} + 1 \right)^{2} A^{2} B^{2} \, .
\end{equation}

Due to the ``cross-diagonal'' form of the local tidal tensor in Equation \ref{eq:tidal_tensor_spherical}, the eigenvalues can be easily found:
\begin{equation}
    \begin{split}
        \lambda_{0} &= 0 \, , \\
        \lambda_{\pm} &= \frac{1}{2} \left[ \tensor{C}{^{1}_{1}} + \tensor{C}{^{3}_{3}} \pm \sqrt{ (\tensor{C}{^{1}_{1}} - \tensor{C}{^{3}_{3}})^{2} + 4 (\tensor{C}{^{1}_{3}})^{2} } \right] \, , \\
        \lambda_{\theta} &= \tensor{C}{^{2}_{2}} \, .
    \end{split}
\end{equation}

It is straightforward to verify that the eigenvector associated with $\lambda_{0}$ points in the time direction $(1, 0, 0, 0)$, while those associated with $\lambda_{\theta}$ point in the polar direction $(0, 0, 1, 0)$ (same as $z$-direction in Cartesian coordinates), and $\lambda_{\pm}$ have mixed components in the radial and azimuthal directions.

Upon substituting the specific metric functions, the eigenvalues correctly reduce to results presented previously in the literature \citep[e.g.,][]{Crispino2016, Sharif2018, Hong2020, Vandeev2021, Vandeev2022, Lima2022, Arora2024}.

It is possible to relax the condition on $\epsilon$ and the angular parts of the metric. The latter is done by setting $g_{\theta\theta} = g_{\phi\phi}/\sin^{2}{\theta} = C(r)$. The tidal tensor still takes the form of Equation \ref{eq:tidal_tensor_spherical}, but the individual components are more complicated. The expressions are too lengthy to be included here, but we invite the interested reader to use the accompanying \texttt{Mathematica} notebook described in Appendix \ref{sec:technical} to derive them.

\section{Technical and numerical details} \label{sec:technical}

\begin{figure}[ht]
    \centering
    \includegraphics[width=\columnwidth]{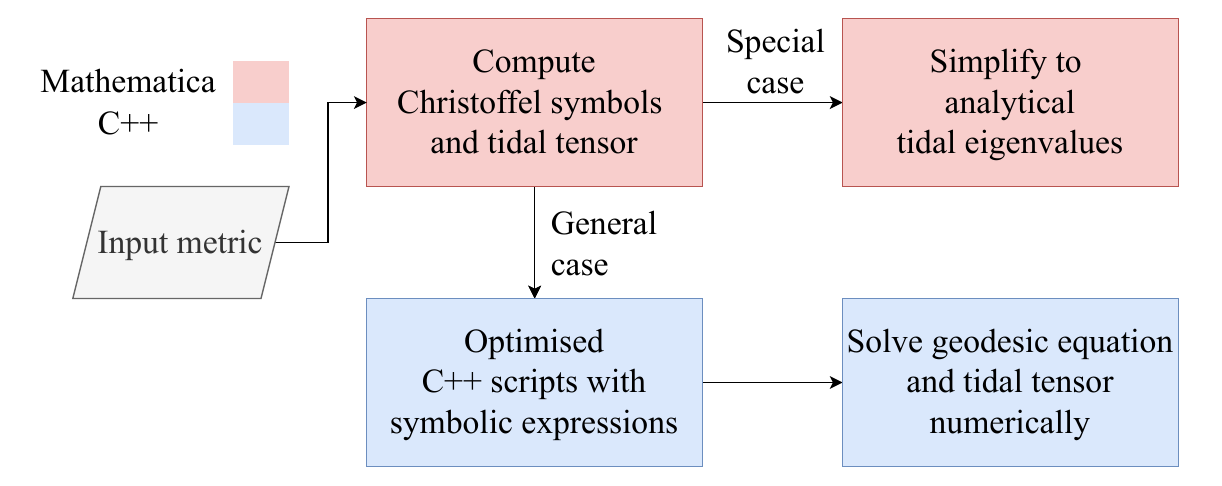}
    \caption{Workflow of our implementation of the ZAMO method.}
    \label{fig:workflow}
\end{figure}

Here we outline the programmatic implementation of the ZAMO method, which is supported by two pieces of software: a \texttt{C++} script for numerical computation and a \texttt{Mathematica} notebook for symbolic manipulation. Both are made available on a public \texttt{GitHub} repository\footnote{\url{https://github.com/Walker-Xin/tidal_zamo}}. The workflow is illustrated in Figure \ref{fig:workflow}. The overall design of the implementation is partly inspired by \texttt{RayTransfer} \cite{Abdikamalov2019,raytransfer2010}

A \texttt{Mathematica} notebook is used to perform all symbolic manipulations given a stationary, axisymmetric metric. The notebook takes in arbitrary metric functions and performs the tensor transformations described in Section \ref{sec:theory}. Simplifications are usually possible when the metric is spherical, but only in rare, specific cases can this be done for an axisymmetric one.

A \texttt{C++} script is written to integrate the geodesic equation and calculate the tidal tensor and its eigenvalues along the trajectory. We implement an eighth-order Runge-Kutta method (``DOP853'') to numerically solve the geodesic equations with adaptive step-size control \cite[Chapter 17]{Press2007}. The \texttt{Eigen} \cite{Eigen2010} library is then used to perform matrix decomposition.

We make use of the coordinate substitution $y = \cos{\theta}$ to transform trigonometric functions into algebraic ones as this significantly improves speed. Given a metric, the \texttt{Mathematica} notebook computes the symbolic expressions of the metric, the Christoffel symbols and the local tidal tensor and outputs them in a separate \texttt{C++} file \cite{Skrivan2019} to be called by the integrator. The use of symbolic expressions also allows for easy modification of the metric.

\bibliography{tidal_disruption, andy}

\end{document}